\documentclass[aps,twocolumn,showpacs,preprintnumbers,floats]{revtex4}\def\@cite#1#2{\textsuperscript{[{#1\if@tempswa , #2\fi}]}}

\usepackage{graphicx}
\usepackage{amsmath}
\usepackage{amsfonts}
\usepackage{amssymb}
\usepackage{color}
\usepackage{subfigure}
\usepackage{epsfig}
\usepackage{morefloats}
\usepackage{multirow}
\usepackage{graphicx,booktabs}
\usepackage{mathrsfs}
\usepackage{txfonts}
\usepackage{indentfirst}
\usepackage{graphicx,booktabs}
\usepackage{epstopdf}
\usepackage{longtable,lscape}
\usepackage[colorlinks, citecolor=blue,anchorcolor=red,menucolor=red, linkcolor=red,filecolor=red,urlcolor=blue,frenchlinks=red]{hyperref}

\newcommand{\vlab}{\mbox{\boldmath$\lambda$\unboldmath}}

\newcommand{\vxi}{\mbox{\boldmath$\xi$\unboldmath}}

\usepackage{comment}

\begin{document}

%\begin{spacing}{2.0}

\title{Fully-heavy tetraquark states and their evidences in the LHC observations }

\author{Ming-Sheng Liu$^{1,4}$, Feng-Xiao Liu$^{1,5}$~\footnote{Feng-Xiao Liu and Ming-Sheng Liu contributed equally to this work.}
Xian-Hui Zhong$^{1,5}$~\footnote {E-mail: zhongxh@hunnu.edu.cn}, Qiang Zhao$^{2,3,5}$~\footnote {E-mail: zhaoq@ihep.ac.cn}}

\affiliation{ 1) Department of Physics, Hunan Normal University, and Key Laboratory of Low-Dimensional Quantum Structures and Quantum Control of Ministry of Education, Changsha 410081, China }

\affiliation{ 2) Institute of High Energy Physics, Chinese Academy of Sciences, Beijing 100049, China}

\affiliation{ 3) University of Chinese Academy of Sciences, Beijing 100049, China}

\affiliation{ 4) College of Science, Tianjin University of Technology, Tianjin 300384, China}

\affiliation{ 5)  Synergetic Innovation Center for Quantum Effects and Applications (SICQEA),
Hunan Normal University, Changsha 410081, China}
%

%\date{\today}

\begin{abstract}

Stimulated by the exciting progress on the observations of the fully-charmed tetraquarks at LHC,
we carry out a combined analysis of the mass spectra and fall-apart decays of the $1S$-, $2S$-, and $1P$-wave $cc\bar{c}\bar{c}$ states in a nonrelativistic quark model (NRQM). It is found that the $X(6600)$ structure observed in the di-$J/\psi$ invariant mass spectrum can be explained by the $1S$-wave state $T_{(4c)0^{++}}(6550)$. This structure may also bear some feed-down effects from the higher $2S$ and/or $1P$ tetraquark states. The $X(6900)$ structure observed in both the di-$J/\psi$
and $J/\psi \psi(2S)$ channels can be naturally explained by the $2S$-wave state $T_{(4c)0^{++}}(6957)$.
The small shoulder structure around $6.2-6.4$ GeV observed at CMS and ATLAS may be due to the feed-down effects from some $1P$-wave states with $C=-1$ and/or some $2S$-wave states with $J^{PC}=0^{++}$. Other decay channels are implied in such a scenario and they can be investigated by future experimental analyses. Considering the large discovery potential at LHC, we also present predictions for the $bb\bar{b}\bar{b}$ states which can be searched for in the future.

\end{abstract}


\maketitle

\section{Introduction}

%{\emph{Introduction}}--
Searching for genuine exotic hadrons beyond the conventional quark model has been one of the most important initiatives since the establishment of the nonrelativistic constituent quark model in 1964~\cite{GellMann:1964nj,Zweig:1981pd}.
Benefited from great progresses in experiment, many candidates of exotic hadrons have been found since the discovery of $X(3872)$ by
Belle in 2003~\cite{Choi:2003ue}. Recent reviews of the status of experimental and theoretical studies can be found in Refs.~\cite{Liu:2019zoy,Esposito:2016noz,Olsen:2017bmm,Lebed:2016hpi,Chen:2016qju,Ali:2017jda,Guo:2017jvc}.
While many observed candidates have been found located in the vicinity of $S$-wave open thresholds, no signals for overall-color-singlet multiquark states have been indisputably established due to difficulties of distinguishing them from hadronic molecules~\cite{Guo:2017jvc}. Recently, the tetraquarks of all-heavy systems, such as $cc\bar{c}\bar{c}$ and $bb\bar{b}\bar{b}$, have received considerable attention. Since the light quark degrees of freedom cannot be exchanged between two heavy mesons at leading order, the color interactions between the heavy quarks (antiquarks) should be dominant at short distance and they may favor to form genuine color-singlet tetraquark configurations rather than loosely bound hadronic molecules. Furthermore, such exotic states may have masses and decay modes significantly different from other conventional states, thus, can be established in experiment.

Early theoretical studies of the full-heavy tetraquark states can be found in the literature~\cite{Ader:1981db,Iwasaki:1975pv,Zouzou:1986qh,Heller:1985cb,Lloyd:2003yc,Barnea:2006sd}. A revival of this topic driven by the experimental progresses can be found by the intensive publications recently~\cite{Wang:2017jtz,Karliner:2016zzc,Berezhnoy:2011xn,Bai:2016int,Anwar:2017toa,Esposito:2018cwh,Chen:2016jxd,
Wu:2016vtq,Hughes:2017xie,Richard:2018yrm,Debastiani:2017msn,Wang:2018poa,Richard:2017vry,Vijande:2009kj,Deng:2020iqw,Ohlsson,
Wang:2019rdo,Bedolla:2019zwg,Chen:2020lgj,Chen:2018cqz,Liu:2019zuc}.
Physicists are very concerned with the stability of the tetraquark $cc\bar{c}\bar{c}$ ($T_{(4c)}$) and $bb\bar{b}\bar{b}$ ($T_{(4b)}$) states.
If the $T_{(4c)}$ or $T_{(4b)}$ states have relatively smaller masses below the thresholds of heavy charmonium or bottomonium pairs~\cite{Wang:2017jtz,Karliner:2016zzc,Berezhnoy:2011xn,Bai:2016int,Anwar:2017toa,Esposito:2018cwh,Debastiani:2017msn,Wang:2018poa}, they may become ``stable" because no direct decays into heavy quarkonium pairs through quark rearrangements would be allowed.
However, some studies showed that stable bound tetraquark states made of $cc\bar{c}\bar{c}$ or $bb\bar{b}\bar{b}$ may not exist~\cite{Wu:2016vtq,Lloyd:2003yc,Ader:1981db,Hughes:2017xie,Richard:2018yrm,Richard:2017vry,Liu:2019zuc,Deng:2020iqw,Wang:2019rdo,Chen:2016jxd,Chen:2018cqz} because the the predicted masses are large enough for them to decay into heavy quarkonium pairs. Due to these very controversial issues, experimental evidence for such exotic objects would be crucial for our understanding of the underlying dynamics.

In 2020, the LHCb Collaboration reported their results on the observations of $T_{(4c)}$ states~\cite{LHCexp}.
In the di-$J/\psi$ invariant mass spectrum, a broad structure above $J/\psi J/\psi$ threshold ranging from 6.2 to 6.8 GeV and a narrower resonance $X(6900)$ were observed with more than 5 $\sigma$ of significance level. There are also some vague structures around 7.2 GeV to be confirmed.
Later in 2022, $X(6900)$ was confirmed in the same final state by both the ATLAS~\cite{ATLASexp} and CMS~\cite{CMSexp} collaborations.
Some signal of $X(6900)$ was also seen in the $J/\psi \psi(2S)$ channel by the ATLAS Collaboration~\cite{ATLASexp}.
In addition, in the lower mass region the CMS measurements show that a clear resonance $X(6600)$
together with a small shoulder structure around $6.2-6.4$ GeV lies in the di-$J/\psi$ spectrum~\cite{CMSexp}.
These clear structures may be evidences for genuine tetraquark $T_{(4c)}$ states, they can also set up experimental constraints on theoretical models of which the successful interpretations and most importantly the early predictions should bring a lot of insights to the underlying dynamics.
Stimulated by the newly observed structures in the di-$J/\psi$ invariant mass spectrum,
the study of full-heavy tetraquark states has been a hot topic in the last two years~\cite{Dong:2022sef,Chen:2022mcr,Faustov:2022mvs,Niu:2022cug,Wang:2022yes,An:2022qpt,Biloshytskyi:2022dmo,Wang:2022jmb,
Gong:2022hgd,Wu:2022qwd,Zhuang:2021pci,Yan:2021glh,Wang:2021mma,Majarshin:2021hex,Wang:2021kfv,
Yang:2021hrb,Huang:2021vtb,Goncalves:2021ytq,Liu:2020tqy,Wang:2020tpt,Wan:2020fsk,Gong:2020bmg,Cao:2020gul,
Guo:2020pvt,Zhu:2020xni,Zhang:2020xtb,Feng:2020riv,Ma:2020kwb,Dong:2020nwy,Wang:2020dlo,Karliner:2020dta,
Wang:2020wrp,Giron:2020wpx,Wang:2020gmd,Chen:2020xwe,Wang:2020ols,Yang:2020rih,Niu:2022vqp,Liang:2022rew,
Liang:2021fzr,Dong:2021lkh,Li:2021ygk,Ke:2021iyh,Yang:2020wkh,Weng:2020jao,Zhao:2020nwy,Zhou:2022xpd,Kuang:2023vac,Zhao:2020jvl,
Becchi:2020mjz,Becchi:2020uvq,Chen:2022sbf}.

In Refs.~\cite{Liu:2019zuc,Liu:2021rtn} we adopted a nonrelativistic potential quark model (NRPQM), which is based on the Hamiltonian proposed by the Cornell model~\cite{Eichten:1978tg}, for the study of fully-heavy tetraquark system. The masses of the $1S$-wave fully-heavy tetraquark states were predicted there and we found that the $1S$-wave $T_{(4c)}$ masses should be above the two-charmonium thresholds within a commonly accepted parameter space~\cite{Liu:2019zuc}. This turns out to be consistent with the structure $X(6600)$.
The predicted masses of the $2S$-wave $T_{(4c)}$ states are comparable with the narrow structure $X(6900)$.
Later studies by Refs.~\cite{Wang:2019rdo,Wang:2021kfv,Lu:2020cns,Li:2021ygk,Zhao:2020jvl} turn out to agree with our predictions.

In this work we carry out a systematic study of the fall-apart decays of the $1S$-, $2S$- and $1P$-wave $T_{(4Q)}$ states in the  NRPQM  framework. The $1S$-, $2S$- and $1P$-wave $T_{(4Q)}$ ($Q=c,b$) states are calculated in the same framework. Thus, it allows us to obtain a self-consistent treatment for the mass spectrum and decay properties. We will show that some of these structures observed at LHC may arise from the $S$- and $P$-wave $T_{(4c)}$ states. To proceed,  we first give a brief introduction to our model and method.

\begin{figure}
\centering \epsfxsize=8.6 cm \epsfbox{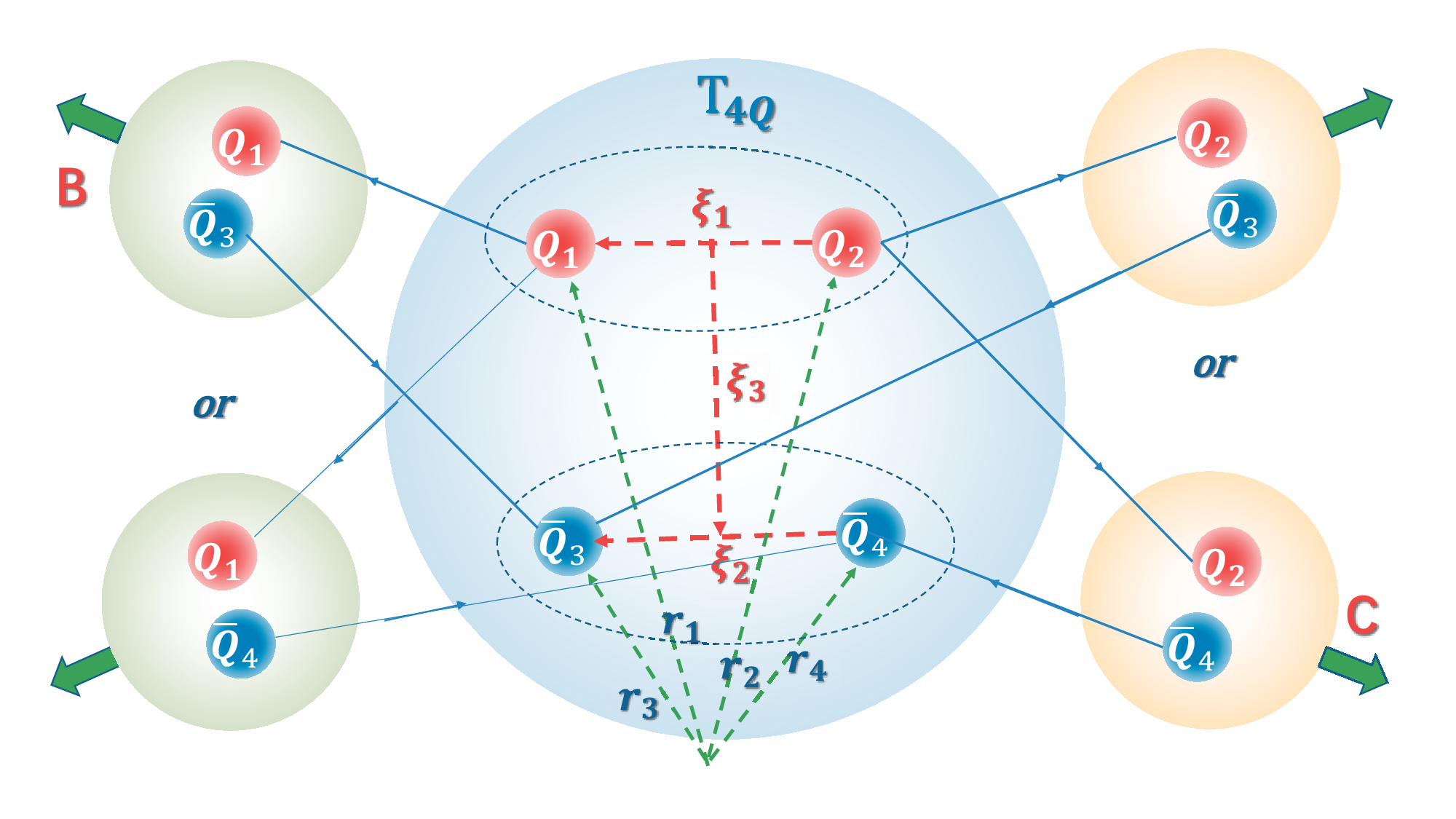}
\caption{The coordinates defined for a $T_{(4Q)}$ system and its fall-apart decays into a $BC$ meson pair via the quark rearrangement.
The $BC$ final state can be formed via two quark rearrangement ways: $(Q_1\bar{Q}_3)(Q_2\bar{Q}_4)$ and $(Q_1\bar{Q}_4)(Q_2\bar{Q}_3)$ as shown in the figure.}
\label{aaa}
\end{figure}

\begin{figure*}[]
\centering \epsfxsize=16.8 cm \epsfbox{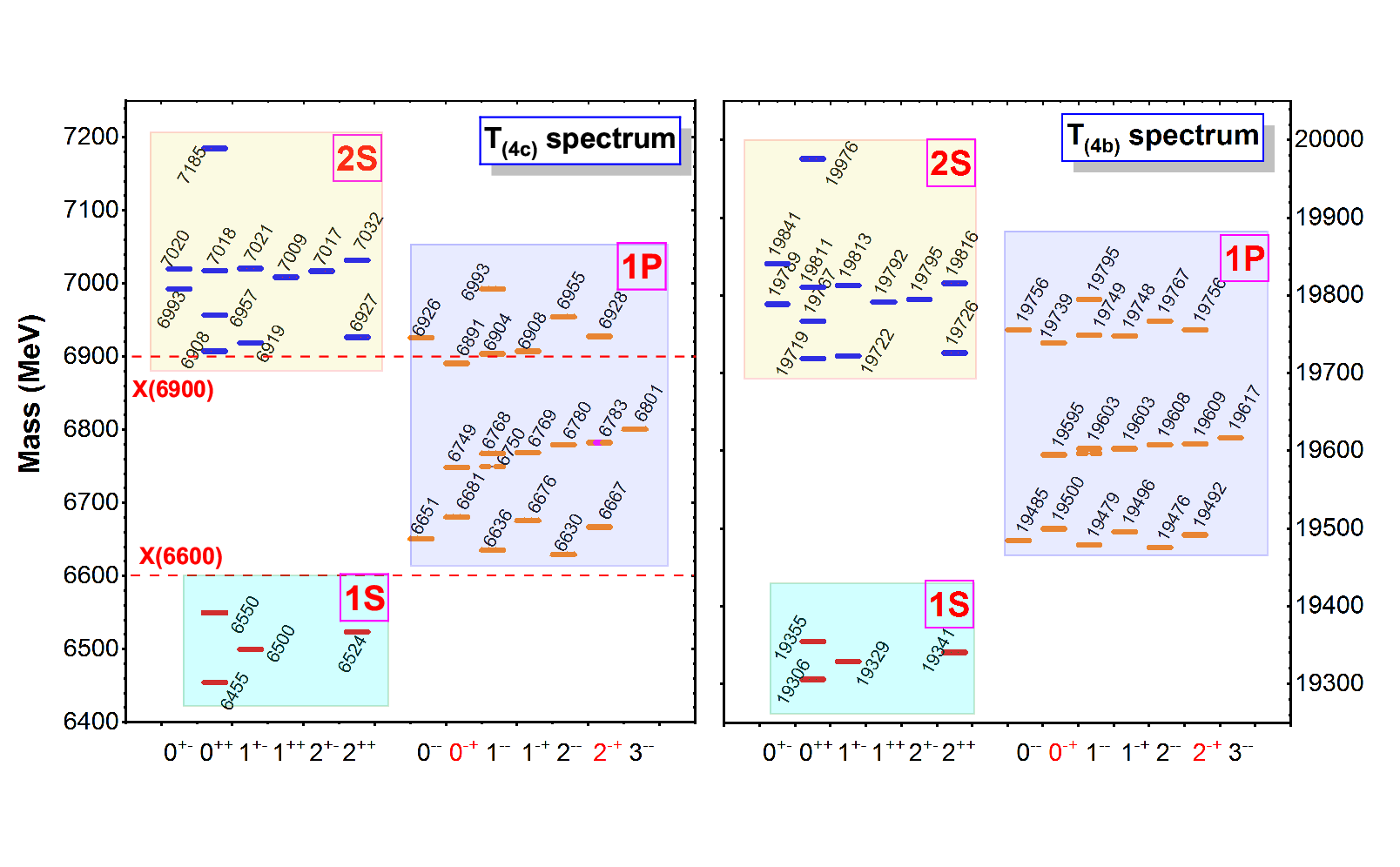}\vspace{-0.9cm} \caption{Mass spectra for the $cc\bar{c}\bar{c}$ and $bb\bar{b}\bar{b}$ systems.}\label{figmasss}
\end{figure*}

\section{Model and method}

%{\emph{Model and method}}--

Apart from the linear confinement, Coulomb type potential, and spin-spin interaction potential for calculating the $S$-wave tetraquark states in the Hamiltonian~\cite{Liu:2019zuc}, we include the spin-orbit and tensor potentials here to deal with the first orbital ($1P$) excitation,
\begin{equation}\label{voge ls}
\begin{split}
V^{LS}_{ij}=&-\frac{\alpha_{ij}}{16}\frac{{\vlab}_i\cdot{\vlab}_j}{r_{ij}^3} \bigg(\frac{1}{m_i^2}+\frac{1}{m_j^2}+\frac{4}{m_im_j}\bigg)\bigg\{\mathbf{L}_{ij}\cdot(\mathbf{S}_{i}+\mathbf{S}_{j})\bigg\}\\
&-\frac{\alpha_{ij}}{16}\frac{{\vlab}_i\cdot{\vlab}_j}{r_{ij}^3}\bigg(\frac{1}{m_i^2}-\frac{1}{m_j^2}\bigg)\bigg\{\mathbf{L}_{ij}\cdot(\mathbf{S}_{i}-\mathbf{S}_{j})\bigg\},
\end{split}
\end{equation}
\begin{equation}\label{voge ten}
V^{T}_{ij}=-\frac{\alpha_{ij}}{4}({\vlab}_i\cdot{\vlab}_j) \frac{1}{m_im_jr_{ij}^3}\Bigg\{\frac{3(\mathbf{S}_i\cdot \mathbf{r}_{ij})(\mathbf{S}_j\cdot \mathbf{r}_{ij})}{r_{ij}^2}-\mathbf{S}_i\cdot\mathbf{S}_j\Bigg\},
\end{equation}
where $r_{ij}\equiv|\mathbf{r}_i-\mathbf{r}_j|$ is the distance between the $i$th and  $j$th quarks, $\mathbf{S}_i$ stands for the spin of the $i$-th quark, and $\mathbf{L}_{ij}$ stands for the relative orbital angular momentum between the $i$-th and $j$-th quark.
If the interaction occurs between two quarks or antiquarks, operator $\vlab_i\cdot\vlab_j$ is defined as
$\vlab_i\cdot\vlab_j\equiv\sum_{a=1}^8\lambda_i^a\lambda_j^a$, while if the interaction occurs between a quark and antiquark, we have $\vlab_i\cdot\vlab_j\equiv\sum_{a=1}^8-\lambda_i^a\lambda_j^{a*}$, where $\lambda^{a*}$ is the complex conjugate of the Gell-Mann matrix $\lambda^a$.
The parameters $b_{ij}$ and $\alpha_{ij}$ denote the confinement potential strength and the strong coupling for the OGE potential, respectively. The same model parameters,  $m_c/m_b=1.483/4.852$ GeV, ${\alpha_{cc}}/{\alpha_{bb}}=0.5461/0.4311$, ${\sigma_{cc}}/{\sigma_{bb}}=1.1384/2.3200$ GeV, and $b_{cc/bb}=0.1425$ GeV$^2$, are adopted by fitting the $c\bar{c}$ and $b\bar{b}$ spectra as in Refs.~\cite{Deng:2016stx,Liu:2019zuc}.

For $T_{(4Q)}$, there are two kinds of color structures, $(6\bar{6})_c$ and $(3\bar{3})_c$. As shown in Fig.~\ref{aaa}, the relative Jacobi coordinate between these two charm quarks (two anticharm quarks) is defined by $\vxi_1=(\textbf{r}_1-\textbf{r}_2)/\sqrt{2}$ ($\vxi_2=(\textbf{r}_3-\textbf{r}_4)/\sqrt{2}$), while the relative Jacobi coordinate between $Q_1Q_2$ and $\bar{Q}_3\bar{Q}_4$
is defined by $\vxi_3=(\textbf{r}_1+\textbf{r}_2)/2-(\textbf{r}_3+\textbf{r}_4)/2$.
Thus, there are three spatial excitation modes which are denoted as $\xi_1$, $\xi_2$, and $\xi_3$.
Their wave functions are defined as $\phi(\xi_i)$ ($i=1,2,3$). According to the requirements of symmetry,
there will be four $1S$ configurations, 12 $2S$ configurations, and 20 $1P$ configurations in the $L-S$ coupling scheme, which are listed in Table~\ref{cccc}. Apart from the conventional quantum numbers, i.e., $J^{PC}=0^{-+}, \ 1^{--}, \ 2^{-\pm}, \ 3^{--}$,
the $P$-wave states can access exotic quantum numbers, i.e., $J^{PC}=0^{--}, \ 1^{-+}$.

To solve the Schr\"{o}dinger equation, we expand the radial part $R_{n_{\xi_i} l_{\xi_i}}(\mathbf{\xi_i})$ of spatial wave function $\phi(\xi_i)$ with a series of harmonic oscillator functions~\cite{Liu:2019vtx}:
\begin{equation}\label{spatial function1}
R_{n_{\xi_i} l_{\xi_i}}(\xi_i)=\sum_{\ell=1}^n\mathcal{C}_{\xi_i\ell}~\phi_{n_{\xi_i} l_{\xi_i}}(d_{\xi_i\ell},\xi_i),
\end{equation}
with
\begin{eqnarray}\label{spatial function2}
%\begin{split}
\phi_{n_{\xi_i}l_{\xi_i}}(d_{\xi_i\ell},\vxi_i)&=\left(\frac{1}{d_{\xi_i\ell}}\right)^{\frac{3}{2}}\Bigg[\frac{2^{l_{\xi_i}+2}}
{(2l_{\xi_i}+1)!!\sqrt{\pi}}\Bigg]^{\frac{1}{2}}\left(\frac{\xi_i}{d_{\xi_i\ell}}\right)^{l_{\xi_i}}e^{-\frac{1}{2}\left(\frac{\xi_i}{d_{\xi_i\ell}}\right)^2}.
%\end{split}
\end{eqnarray}
The parameter $d_{\xi\ell}$ can be related to the harmonic oscillator frequency $\omega_{\xi\ell}$ with $1/d^2_{\xi\ell}=M_\xi\omega_{\xi\ell}$.
For $T_{(4Q)}$ the reduced masses $M_{\xi_i}=m_Q$.
On the other hand, the harmonic oscillator frequency $\omega_{\xi_i\ell}$ can be related to the harmonic oscillator stiffness factor $K_{\ell}$ with $\omega_{\xi_i\ell}=\sqrt{3K_\ell/M_{\xi_i}}$. Then, one has $d_{\xi_i\ell}=d_{\ell}=(3m_Q K_\ell)^{-1/4}$.
The oscillator length $d_\ell$ is set to be
\begin{equation}\label{geometric progression}
d_\ell=d_1a^{\ell-1}\ \ \ (\ell=1,...,n),
\end{equation}
where $n$ is the number of harmonic oscillator functions, and $a$ is the ratio coefficient. There are three parameters $\{d_1,d_n,n\}$ to be determined through the variation method. It is found that with the parameter sets \{0.068 fm, 2.711 fm, 15\} and \{0.050 fm, 2.016 fm, 15\} for the $cc\bar{c}\bar{c}$ and $bb\bar{b}\bar{b}$ systems, we can obtain stable solutions.

\begin{figure}[htbp]
\centering%
\includegraphics[width=0.90\columnwidth]{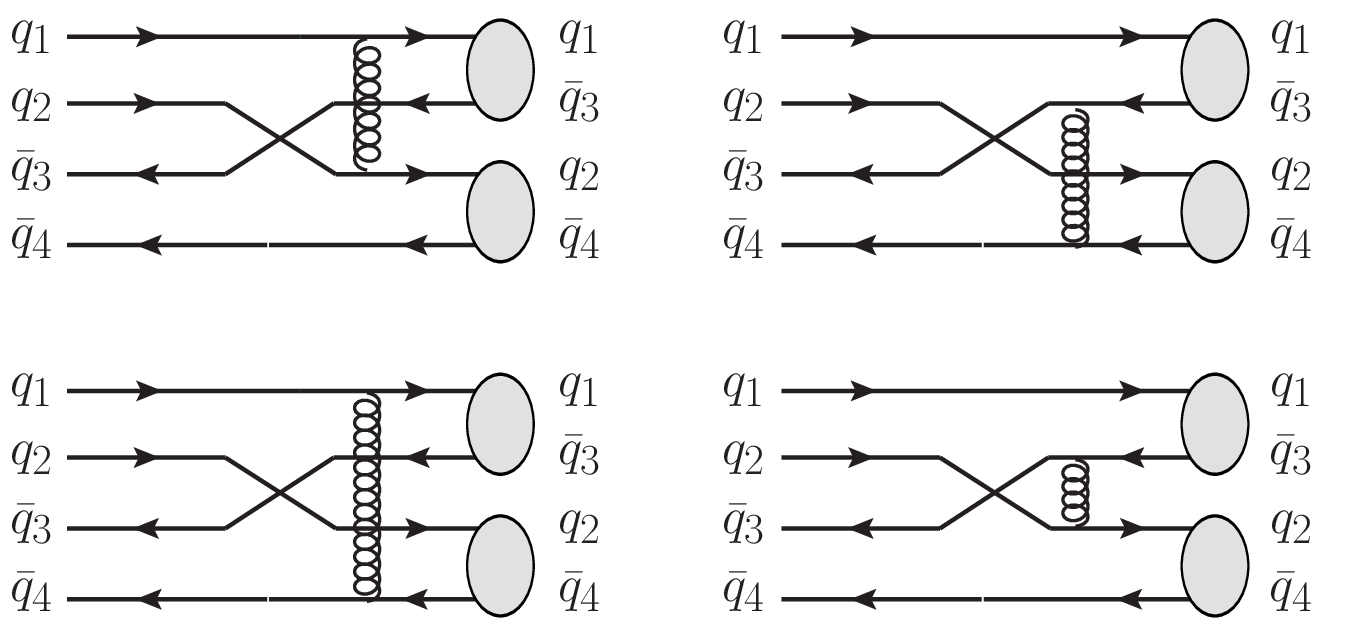}
% \vspace{0.4cm}
\caption{The fall-apart decays of
a $T_{(4Q)}$ state induced by the interactions
$V_{ij}$ ($ij\neq 13,24$ or $ij\neq 14,23$) between inner quarks of final hadrons $B$ and $C$.}
\label{fig:decay}
\end{figure}

By using the spectrum obtained from NRPQM, we further evaluate the fall-apart decays of the $T_{(4Q)}$ states in a
quark-exchange model~\cite{Barnes:2000hu}. The interactions
$V_{ij}$ between inner quarks of final hadrons $B$ and $C$ may be the sources of the fall-apart decays of
a $T_{(4Q)}$ state via the quark rearrangement.
The decay amplitude $\mathcal{M}(A\to BC)$ of $T_{(4Q)}$ state is
described by:
\begin{equation}
\mathcal{M}(A\to BC)=-\sqrt{(2\pi)^{3}}\sqrt{8M_{A}E_{B}E_{C}}\left\langle BC|\underset{i<j}{\sum}V_{ij}|A\right\rangle ,
\end{equation}
where $A$ stands for the initial tetraquark state, $BC$ stands for the final hadron pair. $M_{A}$ is the mass of the initial state, and $E_{B}$ and $E_{C}$ are the energies of the final states $B$ and $C$, respectively.
The decay width $\Gamma$ of $A\to BC$ can be described by:
\begin{eqnarray}
\Gamma & = & \frac{1}{2J_{A}+1}\frac{|\boldsymbol{p}|}{8\pi M_{A}^{2}}\left|\mathcal{M}(A\to BC)\right|^{2},
\end{eqnarray}
where $|\boldsymbol{p}|$ is magnitude of the momentum for the final states $B$ and $C$.
The potentials $V_{ij}$ ($ij\neq 13,24$ or $ij\neq 14,23$) between inner quarks of final hadrons $B$ and $C$, as shown in Fig.~\ref{fig:decay}, are taken the same as our mass calculations. The calculation of the decay amplitude for a
$T_{(4Q)}$ state is indeed a tedious task, some details are given in the appendix.
This model has been developed and applied to the study of the hidden-charm decay properties for the multiquark states in the literature~\cite{Wang:2019spc,Xiao:2019spy,Wang:2020prk,Han:2022fup,Liu:2022hbk}, and a lot of inspiring results are obtained.
For simplicity, the wave functions of the $A,B,C$ hadron states are parametrized out in a single
harmonic oscillator form by fitting the wave functions calculated from our potential model~\cite{Liu:2019vtx,Liu:2020lpw,Liu:2021rtn}.
The harmonic oscillator parameters for the final meson states and initial tetraquark states are collected in
Tables~\ref{mesonp} and ~\ref{t4q} of the appendix, respectively.

\begin{table*}[htp]
\begin{center}
\caption{\label{cccc} Mass spectra for the $1S,1P,2S$-wave $cc\bar{c}\bar{c}$ and $bb\bar{b}\bar{b}$ states. $\xi_1,\xi_2,\xi_3$ are the Jacobi coordinates. $(\xi_1,\xi_2)$ stands for a configuration containing both $\xi_1$- and $\xi_2$-mode orbital excitations, while ($\xi_3$) stands for
a configuration containing $\xi_3$-mode orbital excitation. The spectra for the $1S$ and $2S$ states are taken from our previous works~\cite{Liu:2019zuc,Liu:2021rtn}.   }
\begin{tabular}{ccccc}
\hline \hline
\multirow{1}{*}{Configuration } & \multicolumn{2}{c}{$\underline{~~~~~~~~~~~~~~~~~~~~~~~~~~~~~~~~~~~~cc\bar{c}\bar{c}~~~~~~~~~~~~~~~~~~~~~~~~~~~~~~~~~~~~~~~~}$} & \multicolumn{2}{c}{$\underline{~~~~~~~~~~~~~~~~~~~~~~~~~~~~~~~~~~~~bb\bar{b}\bar{b}~~~~~~~~~~~~~~~~~~~~~~~~~~~~~~~~~~~~~~~~}$}\tabularnewline
\hline
 & Eigenvector  & Mass (MeV)  & Eigenvector  & Mass (MeV) \tabularnewline
\hline
$\begin{array}{l}
1^{1}S_{0^{++}\left(6\bar{6}\right)_{c}}\\
1^{1}S_{0^{++}(\bar{3}3)_{c}}
\end{array}$  & $\left(\begin{array}{rr}
0.58 & 0.81\\
0.81 & -0.58
\end{array}\right)$  & $\left(\begin{array}{c}
6455\\
6550
\end{array}\right)$  & $\left(\begin{array}{rr}
0.58 & 0.81\\
0.81 & -0.58
\end{array}\right)$  & $\left(\begin{array}{c}
19306\\
19355
\end{array}\right)$ \tabularnewline
$\begin{array}{l}
1^{3}S_{1^{+-}(\bar{3}3)_{c}}\end{array}$  & 1  & $6500$  & 1  & $19329$ \tabularnewline
$\begin{array}{l}
1^{5}S_{2^{++}(\bar{3}3)_{c}}\end{array}$  & 1  & $6524$  & 1  & $19341$ \tabularnewline
\hline
$\begin{array}{l}
^{3}P_{0^{--}(6\bar{6})_{c}\left(\xi_{1},\xi_{2}\right)}\\
^{3}P_{0^{--}(\bar{3}3)_{c}\left(\xi_{1},\xi_{2}\right)}
\end{array}$  & $\left(\begin{array}{rr}
-0.80 & 0.60\\
0.60 & 0.80
\end{array}\right)$  & $\left(\begin{array}{c}
6651\\
6926
\end{array}\right)$  & $\left(\begin{array}{rr}
-0.81 & 0.59\\
0.59 & 0.81
\end{array}\right)$  & $\left(\begin{array}{c}
19485\\
19756
\end{array}\right)$ \tabularnewline
$\begin{array}{l}
^{3}P_{0^{-+}(6\bar{6})_{c}\left(\xi_{1},\xi_{2}\right)}\\
^{3}P_{0^{-+}(\bar{3}3)_{c}\left(\xi_{1},\xi_{2}\right)}\\
^{3}P_{0^{-+}(\bar{3}3)_{c}\left(\xi_{3}\right)}
\end{array}$  & $\left(\begin{array}{rrr}
0.82 & 0.47 & 0.32\\
0.14 & 0.38 & -0.91\\
-0.55 & 0.80 & 0.25
\end{array}\right)$  & $\left(\begin{array}{c}
6681\\
6749\\
6891
\end{array}\right)$  & $\left(\begin{array}{rrr}
0.82 & 0.55 & 0.12\\
0.02 & 0.18 & -0.98\\
-0.57 & 0.81 & 0.14
\end{array}\right)$  & $\left(\begin{array}{c}
19500\\
19595\\
19739
\end{array}\right)$ \tabularnewline
$\begin{array}{l}
^{3}P_{1^{--}(6\bar{6})_{c}\left(\xi_{1},\xi_{2}\right)}\\
^{3}P_{1^{--}(\bar{3}3)_{c}\left(\xi_{1},\xi_{2}\right)}\\
^{5}P_{1^{--}(\bar{3}3)_{c}\left(\xi_{3}\right)}\\
^{1}P_{1^{--}(\bar{3}3)_{c}\left(\xi_{3}\right)}\\
^{1}P_{1^{--}(6\bar{6})_{c}\left(\xi_{3}\right)}
\end{array}$  & $\left(\begin{array}{rrrrr}
-0.82 & 0.55 & 0.12 & -0.06 & 0.03\\
0.02 & -0.24 & 0.96 & -0.16 & 0.06\\
-0.01 & 0.05 & 0.17 & 0.98 & 0.10\\
0.48 & 0.69 & 0.19 & -0.02 & -0.50\\
0.31 & 0.39 & 0.02 & -0.11 & 0.86
\end{array}\right)$  & $\left(\begin{array}{c}
6636\\
6750\\
6768\\
6904\\
6993
\end{array}\right)$  & $\left(\begin{array}{rrrrr}
-0.82 & 0.57 & 0.06 & -0.03 & 0.01\\
0.00 & -0.10 & 0.98 & -0.18 & 0.02\\
0.00 & 0.03 & 0.18 & 0.98 & 0.04\\
0.53 & 0.76 & 0.08 & -0.02 & -0.36\\
0.22 & 0.28 & 0.00 & -0.05 & 0.93
\end{array}\right)$  & $\left(\begin{array}{c}
19479\\
19597\\
19603\\
19749\\
19795
\end{array}\right)$ \tabularnewline
$\begin{array}{l}
^{3}P_{1^{-+}(6\bar{6})_{c}\left(\xi_{1},\xi_{2}\right)}\\
^{3}P_{1^{-+}(\bar{3}3)_{c}\left(\xi_{1},\xi_{2}\right)}\\
^{3}P_{1^{-+}(\bar{3}3)_{c}\left(\xi_{3}\right)}
\end{array}$  & $\left(\begin{array}{rrr}
0.82 & 0.56 & 0.05\\
0.01 & 0.08 & -1.00\\
-0.57 & 0.82 & 0.06
\end{array}\right)$  & $\left(\begin{array}{c}
6676\\
6769\\
6908
\end{array}\right)$  & $\left(\begin{array}{rrr}
0.82 & 0.57 & 0.05\\
0.00 & 0.08 & -1.00\\
-0.57 & 0.82 & 0.06
\end{array}\right)$  & $\left(\begin{array}{c}
19496\\
19603\\
19748
\end{array}\right)$ \tabularnewline
$\begin{array}{l}
^{3}P_{2^{--}(6\bar{6})_{c}\left(\xi_{1},\xi_{2}\right)}\\
^{3}P_{2^{--}(\bar{3}3)_{c}\left(\xi_{1},\xi_{2}\right)}\\
^{5}P_{2^{--}(\bar{3}3)_{c}\left(\xi_{3}\right)}
\end{array}$  & $\left(\begin{array}{rrr}
0.80 & -0.59 & -0.06\\
0.01 & 0.12 & -0.99\\
0.60 & 0.80 & 0.10
\end{array}\right)$  & $\left(\begin{array}{c}
6630\\
6780\\
6955
\end{array}\right)$  & $\left(\begin{array}{rrr}
0.81 & -0.59 & -0.03\\
0.00 & 0.06 & -1.00\\
0.59 & 0.81 & 0.05
\end{array}\right)$  & $\left(\begin{array}{c}
19476\\
19608\\
19767
\end{array}\right)$ \tabularnewline
$\begin{array}{l}
^{3}P_{2^{-+}(6\bar{6})_{c}\left(\xi_{1},\xi_{2}\right)}\\
^{3}P_{2^{-+}(\bar{3}3)_{c}\left(\xi_{1},\xi_{2}\right)}\\
^{3}P_{2^{-+}(\bar{3}3)_{c}\left(\xi_{3}\right)}
\end{array}$  & $\left(\begin{array}{rrr}
0.82 & 0.57 & -0.06\\
0.00 & -0.10 & -1.00\\
-0.58 & 0.81 & -0.08
\end{array}\right)$  & $\left(\begin{array}{c}
6667\\
6783\\
6928
\end{array}\right)$  & $\left(\begin{array}{rrr}
0.82 & 0.58 & 0.00\\
0.00 & 0.00 & -1.00\\
-0.58 & 0.82 & 0.00
\end{array}\right)$  & $\left(\begin{array}{c}
19492\\
19609\\
19756
\end{array}\right)$ \tabularnewline
$\begin{array}{c}
^{5}P_{3^{--}(\bar{3}3)_{c}\left(\xi_{3}\right)}\end{array}$  & $1$  & $6801$  & $1$  & $19617$ \tabularnewline
\hline
$\begin{array}{l}
2^{1}S_{0^{+-}\left(6\bar{6}\right)_{c}\left(\xi_{1},\xi_{2}\right)}\\
2^{1}S_{0^{+-}(\bar{3}3)_{c}\left(\xi_{1},\xi_{2}\right)}
\end{array}$  & $\left(\begin{array}{rr}
-0.66 & -0.75\\
-0.75 & 0.66
\end{array}\right)$  & $\left(\begin{array}{c}
6993\\
7020
\end{array}\right)$  & $\left(\begin{array}{rr}
0.10 & -1.00\\
-1.00 & -0.10
\end{array}\right)$  & $\left(\begin{array}{c}
19789\\
19841
\end{array}\right)$ \tabularnewline
$\begin{array}{l}
2^{1}S_{0^{++}\left(6\bar{6}\right)_{c}\left(\xi_{1},\xi_{2}\right)}\\
2^{1}S_{0^{++}(\bar{3}3)_{c}\left(\xi_{1},\xi_{2}\right)}\\
2^{1}S_{0^{++}\left(6\bar{6}\right)_{c}\left(\xi_{3}\right)}\\
2^{1}S_{0^{++}(\bar{3}3)_{c}\left(\xi_{3}\right)}
\end{array}$  & $\left(\begin{array}{rrrr}
0.35 & 0.40 & 0.03 & 0.84\\
-0.91 & -0.05 & 0.07 & 0.41\\
0.21 & -0.91 & -0.01 & 0.35\\
-0.06 & 0.02 & -1.00 & 0.05
\end{array}\right)$  & $\left(\begin{array}{c}
6908\\
6957\\
7018\\
7185
\end{array}\right)$  & $\left(\begin{array}{rrrr}
0.15 & 0.38 & 0.04 & 0.91\\
-0.97 & -0.04 & -0.14 & 0.19\\
0.10 & -0.92 & 0.02 & 0.37\\
0.14 & 0.00 & -0.99 & 0.01
\end{array}\right)$  & $\left(\begin{array}{c}
19719\\
19767\\
19811\\
19976
\end{array}\right)$ \tabularnewline
$\begin{array}{l}
2^{3}S_{1^{+-}(\bar{3}3)_{c}\left(\xi_{1},\xi_{2}\right)}\\
2^{3}S_{1^{+-}(\bar{3}3)_{c}\left(\xi_{3}\right)}
\end{array}$  & $\left(\begin{array}{rr}
-0.39 & -0.92\\
-0.92 & 0.39
\end{array}\right)$  & $\left(\begin{array}{c}
6919\\
7021
\end{array}\right)$  & $\left(\begin{array}{rr}
-0.38 & -0.92\\
-0.92 & 0.38
\end{array}\right)$  & $\left(\begin{array}{c}
19722\\
19813
\end{array}\right)$ \tabularnewline
$\begin{array}{l}
2^{3}S_{1^{++}(\bar{3}3)_{c}\left(\xi_{1},\xi_{2}\right)}\end{array}$  & 1  & $7009$  & 1  & $19792$ \tabularnewline
$\begin{array}{l}
2^{5}S_{2^{+-}(\bar{3}3)_{c}\left(\xi_{1},\xi_{2}\right)}\end{array}$  & 1  & $7017$  & 1  & $19795$ \tabularnewline
$\begin{array}{l}
2^{5}S_{2^{++}(\bar{3}3)_{c}\left(\xi_{1},\xi_{2}\right)}\\
2^{5}S_{2^{++}(\bar{3}3)_{c}\left(\xi_{3}\right)}
\end{array}$  & $\left(\begin{array}{rr}
-0.37 & -0.93\\
-0.93 & 0.37
\end{array}\right)$  & $\left(\begin{array}{c}
6927\\
7032
\end{array}\right)$  & $\left(\begin{array}{rr}
-0.37 & -0.93\\
-0.93 & 0.37
\end{array}\right)$  & $\left(\begin{array}{c}
19726\\
19816
\end{array}\right)$ \tabularnewline
\hline \hline
\end{tabular}
\end{center}
\end{table*}

%\begin{figure*}[]
%\centering \epsfxsize=15.6 cm \epsfbox{decayQQQQ2.eps}\vspace{-0.2cm} \caption{Fall-apart decay properties for the $cc\bar{c}\bar{c}$ and $bb\bar{b}\bar{b}$ systems.}\label{decays}
%\end{figure*}

\begin{table*}[htp]
\begin{center}
\caption{\label{ccccabc} Fall-apart decay properties for the $T_{(4c)}$ states.}
\begin{tabular}{lcccccccc}																	
\hline\hline																	
State	&	~~~~~~$\eta_{c}\eta_{c}$~~~~~~	&	$J/\psi J/\psi$~~~	&	$\eta_{c}\eta_{c}(2S)$~~~	&	$J/\psi\psi(2S)$~~~	&	 $\chi_{c0}\chi_{c0}$/$\chi_{c0}\chi_{c1}$~~~	 &	$\chi_{c0}\chi_{c2}$/$\chi_{c1}\chi_{c1}$~~~	&	 $\chi_{c1}\chi_{c2}$/$\chi_{c2}\chi_{c2}$/$h_ch_c$	&		 \\
\hline																	
$T_{(4c)0^{++}}(6455)(1S)$	&	$1.45$	&	$0.70$	&	$\cdot\cdot\cdot$	&	$\cdot\cdot\cdot$	&	 $\cdot\cdot\cdot$	&	 $\cdot\cdot\cdot$	&	 $\cdot\cdot\cdot$	&		\\
$T_{(4c)0^{++}}(6550)(1S)$	&	$0.12$	&	$1.78$	&	$\cdot\cdot\cdot$	&	$\cdot\cdot\cdot$	&	 $\cdot\cdot\cdot$	&	 $\cdot\cdot\cdot$	&	 $\cdot\cdot\cdot$	&		\\
$T_{(4c)2^{++}}(6524)(1S)$	&	$\cdot\cdot\cdot$	&	$0.00$	&	$\cdot\cdot\cdot$	&	$\cdot\cdot\cdot$	&	 $\cdot\cdot\cdot$	 &	 $\cdot\cdot\cdot$	&	 $\cdot\cdot\cdot$	&		\\
$T_{(4c)0^{++}}(6908)(2S)$	&	$0.61$	&	$0.12$	&	$0.55$	&	$0.09$	&	 $22.5$/~$\cdot\cdot\cdot$	&	$\cdot\cdot\cdot$	&	 $\cdot\cdot\cdot$	&		 \\
$T_{(4c)0^{++}}(6957)(2S)$	&	$0.01$	&	$4.66$	&	$0.05$	&	$3.17$	&	 $70.8$/~$\cdot\cdot\cdot$	&$\cdot\cdot\cdot$		&	 $\cdot\cdot\cdot$	&		 \\
$T_{(4c)0^{++}}(7018)(2S)$	&	$3.14$	&	$1.87$	&	$0.00$	&	$0.07$	&	 $14.8$/~$\cdot\cdot\cdot$	&$0.00$/~$\cdot\cdot\cdot$		 &	 $\cdot\cdot\cdot$	&		 \\
$T_{(4c)0^{++}}(7185)(2S)$	&	$0.00$	&	$0.48$	&	$0.21$	&	$0.14$	&	$1.14$/~$\cdot\cdot\cdot$	 & $0.05$/~$6.12$	&	 $0.53/18.0/5.79$	 &		\\
$T_{(4c)2^{++}}(6927)(2S)$	&	$\cdot\cdot\cdot$	&	$0.36$	&	$\cdot\cdot\cdot$	&	$0.30$	&	 $0.00$/$1.15$	&	 $\cdot\cdot\cdot$	&	 $\cdot\cdot\cdot$	&		\\
$T_{(4c)2^{++}}(7032)(2S)$	&	$\cdot\cdot\cdot$	&	$7.12$	&	$\cdot\cdot\cdot$	&	$2.83$	&	 $0.23$/$98.0$	&	 $325$/$214$	&	 $\cdot\cdot\cdot$	&		 \\
\hline\hline																	
State 	&	$J/\psi J/\psi$	&	$\eta_{c}\chi_{c0}$	&	$\eta_{c}\chi_{c1}$	&	$\eta_{c}\chi_{c2}$	&	$J/\psi h_{c}$	&	 $J/\psi\psi(2S)$/$\eta_{c}\eta_{c}(2S)$	 &	 $\chi_{c0}\chi_{c0}$/$\chi_{c0}\chi_{c1}$/$\chi_{c0}\chi_{c2}$	&		\\																 \hline																	
$T_{(4c)1^{++}}(7009)(2S)$	&	$\cdot\cdot\cdot$	&	$2.43$	&	$4.57$	&	$4.81$	&	$3.79$	&	$0.00/\cdot\cdot\cdot$	&	 $\cdot\cdot\cdot$/0.00/0.00	&		 \\
$T_{(4c)0^{-+}}(6681)(1P)$	&	$0.04$	&	$1.28$	&	$0.00$	&	$0.03$	&	$0.40$	&	 $\cdot\cdot\cdot$	&	$\cdot\cdot\cdot$	 &		\\
$T_{(4c)0^{-+}}(6749)(1P)$	&	$1.56$	&	$0.00$	&	$0.00$	&	$0.00$	&	$5.33$	&	 $\cdot\cdot\cdot$	&	$\cdot\cdot\cdot$	 &		\\
$T_{(4c)0^{-+}}(6891)(1P)$	&	$0.00$	&	$2.38$	&	$0.00$	&	$0.79$	&	$0.06$	&	$0.00/\cdot\cdot\cdot$	&	 $0.00/\cdot\cdot\cdot/\cdot\cdot\cdot$	&		 \\
$T_{(4c)1^{-+}}(6676)(1P)$	&	$0.09$	&	$0.00$	&	$1.99$	&	$0.03$	&	$1.70$	&	 $\cdot\cdot\cdot/0.00$	&	 $\cdot\cdot\cdot$	&		 \\
$T_{(4c)1^{-+}}(6769)(1P)$	&	$0.21$	&	$0.00$	&	$1.22$	&	$0.06$	&	$5.57$	&	 $\cdot\cdot\cdot/0.00$	&	 $\cdot\cdot\cdot$	&		 \\
$T_{(4c)1^{-+}}(6908)(1P)$	&	$0.00$	&	$0.00$	&	$1.43$	&	$0.06$	&	$0.37$	&	$0.00/0.00$	&	 $0.31/\cdot\cdot\cdot/\cdot\cdot\cdot$	 &		 \\
$T_{(4c)2^{-+}}(6667)(1P)$	&	$0.00$	&	$0.02$	&	$0.03$	&	$0.05$	&	$0.03$	&	$\cdot\cdot\cdot$	 &	$\cdot\cdot\cdot$	 &		\\
$T_{(4c)2^{-+}}(6783)(1P)$	&	$0.01$	&	$0.05$	&	$0.07$	&	$1.10$	&	$2.18$	&	$\cdot\cdot\cdot$	&	 $\cdot\cdot\cdot$	 &		\\
$T_{(4c)2^{-+}}(6928)(1P)$	&	$0.01$	&	$0.05$	&	$0.08$	&	$0.88$	&	$1.98$	&	$0.00/\cdot\cdot\cdot$	&	 $0.00/\cdot\cdot\cdot/\cdot\cdot\cdot$	&		 \\
\hline\hline																	
state 	&	$\eta_{c}J/\psi$	&	$\eta_{c}h_{c}$	&	$\chi_{c0}J/\psi$	&	$\chi_{c1}J/\psi$	&	$\chi_{c2}J/\psi$	&	$\eta_{c}\psi(2S)$/	 $\eta_{c}(2S)J/\psi$	&	$\chi_{c0}h_{c}$	\\
\hline																																		
$T_{(4c)1^{+-}}(6500)(1S)$	&	$0.45$	&	$\cdot\cdot\cdot$	&	$\cdot\cdot\cdot$	&	$\cdot\cdot\cdot$	&	 $\cdot\cdot\cdot$	 &	 $\cdot\cdot\cdot$/	 $\cdot\cdot\cdot$	&	$\cdot\cdot\cdot$	\\
$T_{(4c)0^{+-}}(6993)(2S)$	&	$\cdot\cdot\cdot$	&	$3.16$	&	$3.45$	&	$2.65$	&	$0.54$	&		 $\cdot\cdot\cdot$/	 $\cdot\cdot\cdot$	&	 $\cdot\cdot\cdot$	\\
$T_{(4c)0^{+-}}(7020)(2S)$	&	$\cdot\cdot\cdot$	&	$21.1$	&	$0.17$	&	$0.35$	&	$0.34$	&	 $\cdot\cdot\cdot$/	 $\cdot\cdot\cdot$	&	 $\cdot\cdot\cdot$	\\
$T_{(4c)1^{+-}}(6919)(2S)$	&	$0.05$	&	$\cdot\cdot\cdot$	&	$0.02$	&	$0.04$	&	$0.06$	&	$0.09$/	 $0.68$	 &	 $\cdot\cdot\cdot$	\\
$T_{(4c)1^{+-}}(7021)(2S)$	&	$1.98$	&	$\cdot\cdot\cdot$	&	$0.02$	&	$0.07$	&	$0.12$	&	$0.71$/	 $0.61$	 &	$150$	 \\
$T_{(4c)2^{+-}}(7017)(2S)$	&	$\cdot\cdot\cdot$	&	$\cdot\cdot\cdot$	&	$4.70$	&	$3.75$	&	$0.86$	&	 $\cdot\cdot\cdot$/	 $\cdot\cdot\cdot$	 &$\cdot\cdot\cdot$		\\
\hline																	
$T_{(4c)0^{--}}(6651)(1P)$	&	$0.24$	&	$\cdot\cdot\cdot$	&	$0.00$	&	$6.75$	&	$\cdot\cdot\cdot$	&	 $\cdot\cdot\cdot$/	 $\cdot\cdot\cdot$	 &$\cdot\cdot\cdot$		\\
$T_{(4c)0^{--}}(6926)(1P)$	&	$0.12$	&	$\cdot\cdot\cdot$	&	$0.02$	&	$2.76$	&	$0.19$	&	$0.00$/	 $0.00$	&	 $\cdot\cdot\cdot$	\\
$T_{(4c)1^{--}}(6636)(1P)$	&	$0.04$	&	$0.02$	&	$3.28$	&	$0.97$	&	$\cdot\cdot\cdot$	&	 $\cdot\cdot\cdot$/	 $\cdot\cdot\cdot$	&	 $\cdot\cdot\cdot$	\\
$T_{(4c)1^{--}}(6750)(1P)$	&	$0.17$	&	$0.01$	&	$1.92$	&	$3.99$	&	$1.05$	&	$0.00$/	 $0.00$	&	$\cdot\cdot\cdot$	 \\
$T_{(4c)1^{--}}(6768)(1P)$	&	$0.00$	&	$2.77$	&	$0.08$	&	$0.99$	&	$0.79$	&	$0.00$	/	 $0.00$	&	$\cdot\cdot\cdot$	 \\
$T_{(4c)1^{--}}(6904)(1P)$	&	$0.00$	&	$0.17$	&	$2.96$	&	$0.26$	&	$0.17$	&	$0.00$/	 $0.00$	&	$\cdot\cdot\cdot$	 \\
$T_{(4c)1^{--}}(6993)(1P)$	&	$0.03$	&	$1.91$	&	$0.12$	&	$0.76$	&	$1.49$	&	$0.00$/	 $0.00$	 &	$0.17$	\\
$T_{(4c)2^{--}}(6630)(1P)$	&	$0.06$	&	$0.01$	&	$0.01$	&	$0.02$	&	$\cdot\cdot\cdot$	&	$\cdot\cdot\cdot$/	 $\cdot\cdot\cdot$	&	 $\cdot\cdot\cdot$	\\
$T_{(4c)2^{--}}(6780)(1P)$	&	$0.01$	&	$0.01$	&	$0.01$	&	$3.55$	&	$2.00$	&	$0.00$/	 $0.00$	 &	$\cdot\cdot\cdot$	 \\
$T_{(4c)2^{--}}(6955)(1P)$	&	$0.00$	&	$0.00$	&	$0.11$	&	$1.75$	&	$3.45$	&	$0.00$/	 $0.01$	&	$0.00$	\\
$T_{(4c)3^{--}}(6801)(1P)$	&	$0.00$	&	$0.00$	&	$0.16$	&	$0.27$	&	$11.8$	&	$0.00$/$0.00$	&	 $\cdot\cdot\cdot$	 \\
\hline\hline																	
\end{tabular}																	
\end{center}
\end{table*}

\begin{table*}[htp]
\begin{center}
\caption{\label{bbbbabc} Fall-apart decay properties for the $T_{(4b)}$ states.}
\begin{tabular}{lcccccccc}																	
\hline\hline																	
State	&	~~~~~~$\eta_{b}\eta_{b}$~~~~~~	&	~~$\Upsilon\Upsilon$~~	&	~~$\eta_{b}\eta_{b}(2S)$~~	&	~~$\Upsilon\Upsilon(2S)$~~	&	 $\chi_{b0}\chi_{b0}$/$\chi_{b0}\chi_{b1}$	 &	~~$\chi_{b0}\chi_{b2}$/$\chi_{b1}\chi_{b1}$	&	~~$\chi_{b1}\chi_{b2}$/$\chi_{b2}\chi_{b2}$/$h_bh_b$			 \\
\hline																	
$T_{(4b)0^{++}}(19306)(1S)$	&	0.33	&	0.16	&	$\cdot\cdot\cdot$	&	$\cdot\cdot\cdot$	&	$\cdot\cdot\cdot$	&	 $\cdot\cdot\cdot$	&	 $\cdot\cdot\cdot$			 \\
$T_{(4b)0^{++}}(19355)(1S)$	&	0.02	&	0.38	&	$\cdot\cdot\cdot$	&	$\cdot\cdot\cdot$	&	$\cdot\cdot\cdot$	&	 $\cdot\cdot\cdot$	&	 $\cdot\cdot\cdot$			 \\
$T_{(4b)2^{++}}(19341)(1S)$	&	$\cdot\cdot\cdot$	&	0.00	&	$\cdot\cdot\cdot$	&	$\cdot\cdot\cdot$	&	$\cdot\cdot\cdot$	 &	 $\cdot\cdot\cdot$	&	 $\cdot\cdot\cdot$			 \\
$T_{(4b)0^{++}}(19719)(2S)$	&	$0.11$	&	0.21	&	1.04	&	1.53	&	0.20/$\cdot\cdot\cdot$	&	$\cdot\cdot\cdot$	&	 $\cdot\cdot\cdot$			 \\
$T_{(4b)0^{++}}(19767)(2S)$	&	$0.09$	&	2.29	&	1.00	&	15.4	&	14.2/$\cdot\cdot\cdot$	&	$\cdot\cdot\cdot$	&	 $\cdot\cdot\cdot$			 \\
$T_{(4b)0^{++}}(19811)(2S)$	&	3.80	&	1.89	&	24.0	&	8.30	&	5.58/$\cdot\cdot\cdot$	&	0.01/18.2	&	 0.66/$\cdot\cdot\cdot$/233			 \\
$T_{(4b)0^{++}}(19976)(2S)$	&	$0.07$	&	$0.69$	&	$0.32$	&	2.02	&	1.18/$\cdot\cdot\cdot$	& 0.09/4.27	 &0.67/10.4/5.50				 \\
$T_{(4b)2^{++}}(19726)(2S)$	&	$\cdot\cdot\cdot$	&	0.41	&	$\cdot\cdot\cdot$	&	3.23	&	0.00	&	$\cdot\cdot\cdot$	 &	 $\cdot\cdot\cdot$			 \\
$T_{(4b)2^{++}}(19816)(2S)$	&	$\cdot\cdot\cdot$	&	7.41	&	$\cdot\cdot\cdot$	&	41.8	&	0.03/15.2	&	45.3/34.5	 &75.8/$\cdot\cdot\cdot$/$\cdot\cdot\cdot$				\\
\hline	\hline																
State	&	$\Upsilon\Upsilon$	&	$\eta_{b}\chi_{b0}$	&	$\eta_{b}\chi_{b1}$	&	$\eta_{b}\chi_{b2}$	&	$\Upsilon h_{b}$	&	 $\Upsilon\Upsilon(2S)$/$\eta_{b}\eta_{b}(2S)$	&	~~$\chi_{b0}\chi_{b0}$/$\chi_{b0}\chi_{b1}$/$\chi_{b0}\chi_{b2}$			\\
\hline																																		
$T_{(4b)1^{++}}(19792)(2S)$	&	$\cdot\cdot\cdot$	&	5.41 	&	16.3	&	23.9	&	58.0	&	0.00/$\cdot\cdot\cdot$	&	 $\cdot\cdot\cdot$/0.00/0.00			 \\
$T_{(4b)0^{-+}}(19500)(1P)$	&	0.05	&	1.96	&	0.00	&	0.10	&	1.48	&	0.00/$\cdot\cdot\cdot$	&	 $\cdot\cdot\cdot$			 \\
$T_{(4b)0^{-+}}(19595)(1P)$	&	0.16	&	1.26	&	0.00	&	0.15	&	3.74	&	0.00/$\cdot\cdot\cdot$	&	 $\cdot\cdot\cdot$			 \\
$T_{(4b)0^{-+}}(19739)(1P)$	&	0.01	&	0.99	&	0.00	&	0.27	&	0.07	&	0.00/$\cdot\cdot\cdot$	&	 0.00/$\cdot\cdot\cdot$/	 $\cdot\cdot\cdot$		\\
$T_{(4b)1^{-+}}(19496)(1P)$	&	0.02	&	0.00	&	2.00	&	0.07	&	1.94	&	0.00/0.00	&	$\cdot\cdot\cdot$			 \\
$T_{(4b)1^{-+}}(19603)(1P)$	&	0.03	&	0.00	&	2.14	&	0.16	&	3.75	&	0.00/0.00	&	 $\cdot\cdot\cdot$			 \\
$T_{(4b)1^{-+}}(19748)(1P)$	&	0.00	&	0.00	&	0.59	&	0.05	&	0.25	&	0.00/0.00	&	 0.07/$\cdot\cdot\cdot$/$\cdot\cdot\cdot$ 			 \\
$T_{(4b)2^{-+}}(19492)(1P)$	&	0.00	&	0.00	&	0.01 	&	0.03	&	0.03	&	0.00/$\cdot\cdot\cdot$	&	 $\cdot\cdot\cdot$			 \\
$T_{(4b)2^{-+}}(19609)(1P)$	&	0.00	&	0.04	&	0.08	&	1.17	&	1.66	&	0.00/$\cdot\cdot\cdot$	&	 $\cdot\cdot\cdot$			 \\
$T_{(4b)2^{-+}}(19756)(1P)$	&	0.00	&	0.06	&	0.11	&	1.01	&	1.58	&	0.00/$\cdot\cdot\cdot$	&	 0.00/$\cdot\cdot\cdot$/$\cdot\cdot\cdot$ 			 \\
\hline	\hline																
state	&	$\eta_{b}\Upsilon$	&	$\eta_{b}h_{b}$	&	$\chi_{b0}\Upsilon$	&	$\chi_{b1}\Upsilon$	&	$\chi_{b2}\Upsilon$	&	 $\eta_{b}\Upsilon(2S)$/ $\eta_{b}(2S)\Upsilon$	&	$\chi_{b0}h_{b}$	\\																	
\hline																	
$T_{(4b)1^{+-}}(19329)(1S)$	&	0.10	&	$\cdot\cdot\cdot$	&	$\cdot\cdot\cdot$	&	$\cdot\cdot\cdot$	&	$\cdot\cdot\cdot$	 &	 $\cdot\cdot\cdot$	&	 $\cdot\cdot\cdot$			 \\
$T_{(4b)0^{+-}}(19789)(2S)$	&	$\cdot\cdot\cdot$	&	66.8	&	4.10	&	12.0	&	17.3	&	$\cdot\cdot\cdot$	&	 $\cdot\cdot\cdot$	 		 \\
$T_{(4b)0^{+-}}(19841)(2S)$	&	$\cdot\cdot\cdot$	&	15.9	&	3.42	&	10.5	&	16.0	&	$\cdot\cdot\cdot$	&	 $\cdot\cdot\cdot$	 		 \\
$T_{(4b)1^{+-}}(19722)(2S)$	&	0.20	&	$\cdot\cdot\cdot$	&	0.00	&	0.01	&	0.01	&	267/303	&	$\cdot\cdot\cdot$			 \\
$T_{(4b)1^{+-}}(19813)(2S)$	&	1.82	&	$\cdot\cdot\cdot$	&	0.00	&	0.01	&	0.01	&	82.7/102	&	27		 \\
$T_{(4b)2^{+-}}(19795)(2S)$	&	$\cdot\cdot\cdot$	&	$\cdot\cdot\cdot$	&	12.6	&	37.0	&	54.0	&	$\cdot\cdot\cdot$	 &	 $\cdot\cdot\cdot$	 		 \\
\hline																	
$T_{(4b)0^{--}}(19485)(1P)$	&	0.04	&	$\cdot\cdot\cdot$	&	0.00	&	5.07	&	0.05	&	0.00/0.00	&	 $\cdot\cdot\cdot$			 \\
$T_{(4b)0^{--}}(19756)(1P)$	&	0.02	&	$\cdot\cdot\cdot$	&	0.00	&	0.98	&	0.11	&	0.00/0.00	&	 $\cdot\cdot\cdot$	 		 \\
$T_{(4b)1^{--}}(19479)(1P)$	&	0.01	&	0.00	&	2.01	&	1.09	&	1.90	&	 0.00/0.00	&	$\cdot\cdot\cdot$			 \\
$T_{(4b)1^{--}}(19597)(1P)$	&	0.02	&	0.09	&	3.00	&	2.49	&	0.68	&	 0.00/0.00	&	$\cdot\cdot\cdot$			 \\
$T_{(4b)1^{--}}(19603)(1P)$	&	0.00	&	3.65	&	0.03	&	0.91	&	0.67	&	 0.00/0.00	&	$\cdot\cdot\cdot$			 \\
$T_{(4b)1^{--}}(19748)(1P)$	&	0.00	&	0.02	&	0.89	&	0.21	&	0.13	&	0.00/0.00	&	 $\cdot\cdot\cdot$			 \\
$T_{(4b)1^{--}}(19795)(1P)$	&	0.00	&	0.99	&	0.14	&	0.55	&	0.54	&	0.00/0.00	&	 	0.04 	\\
$T_{(4b)2^{--}}(19476)(1P)$	&	0.01	&	0.00	&	0.00	&	0.01	&	0.06	&	0.00/0.00	&	 $\cdot\cdot\cdot$			 \\
$T_{(4b)2^{--}}(19608)(1P)$	&	0.00	&	0.00	&	0.01	&	2.52	&	1.16	&	0.00/0.00	&	 $\cdot\cdot\cdot$			 \\
$T_{(4b)2^{--}}(19767)(1P)$	&	0.00	&	0.00	&	0.09	&	1.08	&	2.17	&	0.00/0.00	&	0.00 	\\
$T_{(4b)3^{--}}(19617)(1P)$	&	0.00	&	0.00	&	0.15	&	0.30	&	6.85	&	0.00/0.00	&	$\cdot\cdot\cdot$		 \\
\hline\hline																	
\end{tabular}																	
\end{center}
\end{table*}

\section{Results and discussion}

%{\emph{Results and discussion}}--

In Table~\ref{cccc}, the mass spectra of the $T_{(4c)}$ and $T_{(4b)}$ states are listed in the third and fifth columns, respectively.
For clarity, the mass spectra are also plotted in Fig.~\ref{figmasss}.
From Table~\ref{cccc}, one can see that the physical states are usually mixtures of
two different color configurations $|6\bar{6}\rangle_c$ and
$|\bar{3}3\rangle_c$. The eigenvectors for different configurations of the $T_{(4Q)}$ states are also listed in Table~\ref{cccc}.
The eigenvalues for the physical states can be extracted by diagonalizing the mass matrices.
The masses of the $1P$-wave $T_{(4c)}$ and $T_{(4b)}$ states are predicted to be
in the range of $\sim6.6-7.0$ GeV and $\sim19.5-19.8$ GeV, respectively.
The masses of some $1P$-wave $T_{(4c)}$ states are comparable with the newly observed structures $X(6600)$ and $X(6900)$.

It should be mentioned that except for the color configurations $|6\bar{6}\rangle_c$ and
$|\bar{3}3\rangle_c$, one can also select the $|11\rangle_c$ and
$|88\rangle_c$ representations when constructing the tetraquark wave functions.
The two sets of color configurations are equivalent
to each other. The $|6\bar{6}\rangle_c$ and $|\bar{3}3\rangle_c$ configurations can be
expressed by $|11\rangle_c$ and
$|88\rangle_c$ through the Fierz transformation~\cite{Wang:2019rdo}.
With which, one can extracted the $|11\rangle_c$ and
$|88\rangle_c$ components in a physical states expressed with the $|6\bar{6}\rangle_c$ and
$|\bar{3}3\rangle_c$ configurations. The components of different color configurations
for the physical $T_{(4c)}$ and $T_{(4b)}$ states are given in the Tables~\ref{rmsc} and~\ref{rmsb}
of the appendix. To know the spatial size of the tetraquark states,
we also calculate the root mean square radius, our results are also listed in Tables~\ref{rmsc} and~\ref{rmsb}.
Similar to ours, a systematical study of the $1S$, $2S$ and $1P$-wave $T_{(4c)}$
states was also carried out within the non-relativistic quark model in
Ref.~\cite{Wang:2021kfv}. The obtained results are generally
consistent with ours. The slight differences are mainly due
to the different selections of the model parameters, spin-orbital potentials, and numerical methods.
The differences of the numerical methods adopted in the present work and that
in Ref.~\cite{Wang:2021kfv} have been discussed in Ref.~\cite{Wang:2019rdo}.

In addition to calculating the mass spectra, the results of the fall-apart decays
via the quark rearrangement of the $1S$-, $2S$- and $1P$-wave $T_{(4Q)}$ states
are also given in Tables~\ref{ccccabc} and \ref{bbbbabc}. To our surprise, the fall-apart decay widths for
most of the $T_{(4Q)}$ states are only in a small range of $\sim 0-10$ MeV. Thus,
there should exist some stable $T_{(4Q)}$ states although their masses
are above the thresholds of $Q\bar{Q}$ meson pairs.

In the following part, we focus on the $1S$-,
$2S$- and $1P$-wave $T_{(4c)}$ states to understand the di-$J/\psi$ spectrum observed at
LHCb~\cite{LHCexp}, CMS~\cite{CMSexp} and ATLAS~\cite{ATLASexp}.
The nature of the broad structure around $6.2-6.8$ GeV in the di-$J/\psi$ invariant mass spectrum~\cite{LHCexp,CMSexp,ATLASexp} is mysterious if it is a genuine state since it is difficult to understand what decay channels would contribute to its broad width. Note that the CMS~\cite{CMSexp} and ATLAS~\cite{ATLASexp} measurements show some details where a small shoulder appears at the lower side of the broad structure $X(6600)$.

There are no $T_{(4c)}$ states lying in the mass range of $<6.4$ GeV in our NRPQM predictions. However, it turns out to be possible that the small shoulder structure around $6.2-6.4$ GeV near the di-$J/\psi$ threshold
may be caused by some feed-down effects from higher mass $T_{(4c)}$ states.
It is interesting to find that in the $2S$-wave $T_{(4c)}$ states,
the quark rearrangement decay rates of $T_{(4c)0^{++}}(6908,6957,7018)\to \chi_{c0}\chi_{c0}$,
$T_{(4c)0^{++}}(7185)\to \chi_{c1}\chi_{c1},\chi_{c2}\chi_{c2}$
are large, and their partial widths are predicted to be $\mathcal{O}(10)$ MeV.
These $2S$-wave states with $J^{PC}=0^{++}$ have masses around $\sim 6.9-7.2$ GeV. The
$\chi_{cJ} \chi_{cJ'}$ final states can feed down to the di-$J/\psi$ channel via $\chi_{cJ} \chi_{cJ'} \to J/\psi J/\psi+ \gamma\gamma$ and $J/\psi J/\psi+\pi\pi$ where the two soft photons or soft pions  will evade the detection.

We actually find that there could be multi-sources contributing to the shoulder structure around $6.2-6.4$ GeV via the  feed-down mechanisms from some $1P$-wave $T_{(4c)}$ states. For instance, the decay rates for $T_{(4c)0^{--}}(6651)\to J/\psi \chi_{c1}$,
$T_{(4c)1^{--}}(6636)\to J/\psi \chi_{c0}$, $T_{(4c)1^{--}}(6750)\to J/\psi \chi_{c0,1,2}$, $T_{(4c)2^{--}}(6780)\to J/\psi \chi_{c1,2}$, and $T_{(cc\bar{c}\bar{c})3^{--}}(6801)\to J/\psi \chi_{c2}$
are quite large. Their partial widths are predicted to be about several MeV. These $1P$-wave states have masses in the range of $\sim 6.6-6.8$ GeV. Their decays into $J/\psi \chi_{cJ}$ can also contribute to the di-$J/\psi$ channel through the radiative decays of $J/\psi\chi_{c0,1,2}\to J/\psi J/\psi \gamma$, where the photon momenta are about $300\sim 450$ MeV. The feed-down mechanism seems to be a possible explanation for the broad structure around $6.2-6.4$ GeV in the di-$J/\psi$ invariant mass spectrum.

The $X(6600)$ structure observed at CMS may be assigned to the $1S$-wave state $T_{(4c)0^{++}}(6550)$
predicted in the NRPQM. This state is a mixed state between two
color configurations $6\bar{6}$ and $\bar{3}3$. The predicted mass is in good agreement with the observations.
Furthermore, the quark rearrangement decays of $T_{(4c)0^{++}}(6550)$ are governed
by the di-$J/\psi$ channel, which is also consistent with the observations.
Although the predicted partial width of the di-$J/\psi$ mode, $\Gamma_{J/\psi J/\psi}\simeq 1.78$ MeV,
is much smaller than the observed total width $\Gamma=124\pm 29\pm 34$ MeV of $X(6600)$, its width may be saturated by
the hadronic decays into open-charmed meson pairs via the $c\bar{c}$ annihilations~\cite{Anwar:2017toa}.
It was shown in Ref.~\cite{Anwar:2017toa} that the sum of the partial widths of these hadronic decay processes can reach up to order of 100 MeV. In the same picture the other $1S$ state $T_{(4c)0^{++}}(6455)$ may also
be observed in the di-$J/\psi$ channel given the accumulation of more data in the future.
Finally, it should be mentioned that the $X(6600)$ structure may also bear some feed-down
effects from the $2S$-wave states $T_{(4c)0^{+-}}(6978)$ and $T_{(4c)2^{+-}}(7031)$, and/or
the $1P$-wave states $T_{(4c)0^{--}}(6926)$, $T_{(4c)1^{--}}(6904,6993)$ and $T_{(4c)2^{--}}(6955)$,
since they have sizeable partial widths into $J/\psi \chi_{cJ}$ final states.

Concerning the nature of $X(6900)$ the mass location suggests that several $1P$- and $2S$-wave states with $C=+1$ can be the candidates of $X(6900)$. However, with the decay properties taken into account it shows that only the $2S$-wave state $T_{(4c)0^{++}}(6957)$ can match $X(6900)$. The partial widths of $T_{(4c)0^{++}}(6957)$ decaying into the di-$J/\psi$ and $J/\psi \psi(2S)$ channels are predicted to be $\sim4.7$ MeV and $3.2$ MeV, respectively.
Combined with the measured width $\Gamma=122\pm 22\pm 19$ MeV of $X(6900)$, it is predicted that
the decay rates of these two channels are $\mathcal{O}$(3\%).  As listed in Table~\ref{cccc}, $T_{(4c)0^{++}}(6957)$ contains dominantly $2^{1}S_{0^{++}\left(6\bar{6}\right)_{c}\left(\xi_{1},\xi_{2}\right)}$. Its mass slightly higher than  $T_{(4c)0^{++}}(6908)$ of which the dominant configuration is $2^{1}S_{0^{++}(\bar{3}3)_{c}\left(\xi_{3}\right)}$. This is due to the strong attraction produced by the relatively small but crucial mixing of the $2^{1}S_{0^{++}(\bar{3}3)_{c}\left(\xi_{3}\right)}$ configuration. Its crucial role for the $2^{1}S_{0^{++}}$ multiplets can be seen clearly by the mixing matrix in Table~\ref{cccc} for both $cc\bar{c}\bar{c}$ and $bb\bar{b}\bar{b}$.

With $X(6900)$ assigned as the $T_{(4c)0^{++}}(6957)$, the partial width ratio between
the di-$J/\psi$ and $J/\psi \psi(2S)$ channels is predicted to be
\begin{equation}
\frac{\Gamma_{J/\psi J/\psi}}{\Gamma_{J/\psi \psi(2S)}}\simeq 1.5,
\end{equation}
which can be tested in future experiments. As shown in Tab.~\ref{ccccabc}, the main decay channel of $T_{(4c)0^{++}}(6957)$ should be  $T_{(4c)0^{++}}(6957)\to \chi_{c0}\chi_{c0}$. Therefore, a search for $X(6900)$ in the $\chi_{c0}\chi_{c0}$ channel could be useful for understanding its nature. We notice that some other analyses also prefer $X(6900)$ as a compact tetraquark state with $J^{PC}=0^{++}$~\cite{Karliner:2020dta,Zhou:2022xpd,Kuang:2023vac}.
Finally, it should be mentioned that the $2S$ state $T_{(4c)0^{++}}(7018)$ may also contribute to the $X(6900)$ structure observed in
the di-$J/\psi$ final state, since it has a sizeable partial decay width, $\Gamma_{J/\psi J/\psi}\simeq 1.87$ MeV,
into the di-$J/\psi$ channel. Moreover, this state has large decay rates into the $\eta_c\eta_c$
and $\chi_{c0}\chi_{c0}$ channels as well.

\section{Summary}

%{\emph{Summary}}--

With the coherent study of the fully-heavy tetraquark spectra in the NRPQM and their rearrangement decays, we show that the recent measurements of the di-$J/\psi$ spectrum have provided a strong evidence for the $S$- and $P$-wave $T_{(4c)}$ states. The small shoulder structure around 6.2-6.4 GeV observed by CMS and ATLAS
may be due to the feed-down effects from higher $1P$-wave states with $C=-1$
or some $2S$-wave states with $J^{PC}=0^{++}$. The $X(6600)$ structure may arise from the $1S$-wave state
$T_{(4c)0^{++}}(6550)$, of which the peak structure may also bear some feed-down
effects from the $2S$-wave and/or $1P$-wave states with $C=-1$.
The $X(6900)$ structure is most likely to be the $2S$-wave state $T_{(4c)0^{++}}(6957)$.
If $X(6600)$ and $X(6900)$ indeed correspond to the $1S$- and $2S$-wave $T_{(4c)}$ states,
respectively, their decay rates into the di-$J/\psi$ channel are predicted to be order of $1\%$.
In addition, one $1S$-state $T_{(4c)0^{++}}(6455)$, one $2S$-state $T_{(4c)0^{++}}(7016)$ and one
$1P$-state $T_{(4c)0^{-+}}(6749)$ are predicted to be located at the same masses as $X(6600)$ and $X(6900)$ in the di-$J/\psi$ invariant mass spectrum.

Based on such a scenario, we expect that more signals for the $T_{(4c)}$ states should be observed in other decay channels via either $S$- or $P$-wave transitions, such as $\eta_c\eta_c$, $J/\psi h_c$, $J/\psi \chi_{cJ}$, $\eta_c h_c$.  By extending the calculations to the full-bottom tetraquark systems, we have also included the mass spectra of the $bb\bar{b}\bar{b}$ states and their fall-apart decays in Tab.~\ref{cccc} and Tab.~\ref{bbbbabc}, respectively. Several $2S$-wave $T_{(4b)}$ states, such as $T_{(4b)0^{++}}(19767)$ and $T_{(4b)0^{++}}(19811)$, should have good potentials to be observed in di-$\Upsilon$ and $\Upsilon\Upsilon(2S)$ decay channels. Notice that no signals of the $T_{(4b)}$ states are found based on the present statistics at LHCb~\cite{LHCexp}. This can be due to the low production rates for such heavy objects.

\section*{Acknowledgements }

This work is supported by the National Natural Science Foundation of China (Grants No.12175065, No.12235018, No.12105203, No.11775078, and No.U1832173). Q.Z. is also supported in part, by the DFG and NSFC funds to the Sino-German CRC 110 ¡°Symmetries and the Emergence of
Structure in QCD¡± (NSFC Grant No. 12070131001, DFG Project-ID 196253076), National Key Basic Research
Program of China under Contract No. 2020YFA0406300, and Strategic Priority Research Program of Chinese
Academy of Sciences (Grant No. XDB34030302).

\appendix

\section{}

%\newpage

%For a decay process $A\to BC$, in the rest frame of the initial $T_{(4Q)}$ state, the momenta of
%two final meson states $B$ and $C$ are denoted by $\boldsymbol{p_{A}}$ and $\boldsymbol{p_{B}}$, respectively.
%Then one has $\boldsymbol{p}\equiv \boldsymbol{p_{A}}=-\boldsymbol{p_{B}}$. The magnitude of the momentum
%$\boldsymbol{p}$ can be determined by:
%\begin{equation}
%|\boldsymbol{p}|=\frac{\sqrt{\left(M_{A}^{2}-\left(M_{B}+M_{C}\right)^{2}\right)\left(M_{A}^{2}-\left(M_{B}-M_{C}\right)^{2}\right)}}{2M_{A}}.
%\end{equation}

%\subsection{Decay amplitude calculations}

The calculation of the decay amplitude $\mathcal{M}(A\to BC)$ of a
$T_{(4Q)}$ state is indeed a tedious task. For simplicity, the spatial wave functions for the $A,B,C$ hadron states are adopted
a harmonic oscillator form, the harmonic oscillator parameters are determined by fitting the wave functions
calculated from the potential model. These parameters have been given in Tab.~\ref{mesonp} and Tab.~\ref{t4q}.

\begin{table}[htbp]
\centering{}\caption{The masses (MeV) and harmonic oscillator parameters $\beta$ (MeV) for the meson states.}\label{mesonp}.
\scalebox{1.0}{%
\begin{tabular}{l|ccc|ccc}
\hline \hline
State  & $c\bar{c}$  & Mass  & $\beta$  & $b\bar{b}$  & Mass  & $\beta$ \tabularnewline
\hline
$1^{1}S_{0}$  & $\eta_{c}$  & 2984         & 658  & $\eta_{b}$     & 9378      & 1160 \tabularnewline
$1^{3}S_{1}$  & $J/\psi$  & 3097           & 564  & $\varUpsilon$  & 9436      & 1096 \tabularnewline
$2^{1}S_{0}$  & $\eta_{c}$(2S)  & 3635     & 506  & $\eta_{b}$(2S) & 9973      & 858 \tabularnewline
$2^{3}S_{1}$  & $\psi$(2S)  & 3679         & 470  & $\varUpsilon$(2S)  & 9989  & 822 \tabularnewline
$1^{3}P_{0}$  & $\chi_{c0}$  & 3417        & 533  & $\chi_{b0}$  & 9845        & 822 \tabularnewline
$1^{1}P_{1}$  & $h_{c}$  & 3522    & 459  & $h_{b}$  & 9899            & 731 \tabularnewline
$1^{3}P_{1}$  & $\chi_{c1}$ & 3516  & 459  & $\chi_{b1}$  & 9891        & 759 \tabularnewline
$1^{3}P_{2}$  & $\chi_{c2}$  & 3552        & 429  & $\chi_{b2}$  & 9914        & 705 \tabularnewline
\hline \hline
\end{tabular}}
\end{table}

Taking $T_{(cc\bar{c}\bar{c})0^{++}}(1S)(6455)\to J/\psi J/\psi$
as an example, from Tab.~\ref{cccc} one can obtain the wave function
for the initial $T_{(cc\bar{c}\bar{c})0^{++}}(1S)(6455)$ state, i.e.,
\begin{eqnarray}
\left|A\right\rangle  & = & 0.58\psi_{000}^{1S}\chi_{00}^{00}\left|6\bar{6}\right\rangle_c +0.81\psi_{000}^{1S}\chi_{00}^{11}\left|\bar{3}3\right\rangle_c ,\label{eq:WFA}
\end{eqnarray}
which is an admixture between several different configurations. The
wave function of the final state is obtained within the $J$-$J$
coupling scheme. Combining with the Clebsch-Gordan coefficients, the
wave function of the di-$J/\psi$ system with quantum numbers $J^{PC}=0^{++}$
is given by
\begin{eqnarray}
\left|BC\right\rangle  & = & \frac{1}{\sqrt{3}}\left[\chi_{11}^{1}\chi_{1-1}^{2}-\chi_{10}^{1}\chi_{10}^{2}+\chi_{1-1}^{1}\chi_{11}^{2}\right]\nonumber \\
 &  & \varphi_{000}^{1}\varphi_{000}^{2}\phi\left|11\right\rangle , \label{eq:WFBC}
\end{eqnarray}
where the superscripts $1,2$ stand for the two $J/\psi$ mesons in
the final state, $\chi$ stands for the their spin wave functions,
and $\varphi_{000}$ stands for their spatial wave functions. $\phi$
is the wave function for describing the relative motion of two final
state mesons, which is adopted a plane wave form by treating the final
state mesons as free particles:
\begin{eqnarray}
\phi & = & \frac{1}{\left(2\pi\hbar\right)^{\frac{3}{2}}}e^{-i\boldsymbol{p}_{f}.(\boldsymbol{r}_{f_{1}}-\boldsymbol{r}_{f_{2}})},\label{eq:pmb}
\end{eqnarray}
where $\boldsymbol{p}_{f}$ is the three-momentum of the hadron $1$ in the final
state, $\boldsymbol{r}_{f_{1}}$ and $\boldsymbol{r}_{f_{2}}$
stand for the position coordinates of the hadrons $1,2$ in the final
state.

By using the wave functions given in Eqs.~\ref{eq:WFA} and \ref{eq:WFBC}, one can calculate the
transition matrix element with
\begin{eqnarray}
\left\langle BC\right| V_{ij}\left|A\right\rangle & = & \left(\left\langle \frac{1}{\sqrt{3}}\left[\chi_{11}^{1}\chi_{1-1}^{2}-\chi_{10}^{1}\chi_{10}^{2}+\chi_{1-1}^{1}\chi_{11}^{2}\right]\left[11\right]_c\right|\right.\nonumber \\
 &  & \left.\hat{O}_{ij}^{sc}\left|c_{1}\chi_{00}^{00}\left[6\bar{6}\right]+c_{2}\chi_{00}^{11}\left[\bar{3}3\right]_c\right\rangle \right)\nonumber \\
 &  & \left\langle \varphi_{000}^{1}\varphi_{000}^{2}\phi\right|\hat{O}_{ij}^{o}\left|\psi_{000}^{1S}\right\rangle ,
\end{eqnarray}
where the $\hat{O}_{ij}^{sc}$ and $\hat{O}_{ij}^{o}$ stand for the spin-color dependent and spatial dependent operator, respectively.
Calculating the matrix elements in color and spin space is relatively
simple. When calculating the matrix element of the spatial part, $\left\langle \varphi_{000}^{1}\varphi_{000}^{2}\phi\right|\hat{O}_{ij}^{o}\left|\psi_{000}^{1S}\right\rangle$, one should face
a problem. The spatial wave function, which contains three different
variables, cannot be separated into the product of three functions
with independent variables directly. One should solve this problem
by defining new coordinate systems via coordinate transformations
by using the standard linear algebra methods~\cite{Brink:1998as,Fedorov:2017bcq}.

Then, the integration of the spatial part is shown. It is given by
\begin{eqnarray}
 &  & \left\langle \varphi_{000}^{1}\left(\omega_{f_{1}}\right)\varphi_{000}^{2}\left(\omega_{f_{2}}\right)\phi\right|\hat{O}_{ij}^{o}\left|\psi_{000}^{1S}\left(\omega_{i}\right)\right\rangle \nonumber \\
 & = & I_{Nor}\int\hat{O}e^{-\underset{i,j}{\sum}A_{ij}\boldsymbol{\xi}_{i}\cdot\boldsymbol{\xi}_{j}}e^{-i\boldsymbol{p}_{f}.\boldsymbol{\xi}_{3}}\left(Y_{00}\right)^{5}d^{3}\boldsymbol{\xi}_{1}d^{3}\boldsymbol{\xi}_{2}d^{3}\boldsymbol{\xi}_{3},
\end{eqnarray}
where $I_{Nor}$ is a normalization factor independent of the integration
variable, and the matrix $A$ is given by
\begin{equation}
A=\left(\begin{array}{ccc}
\frac{1}{2}\mu\omega_{i}^{2}+\frac{1}{4}\mu\omega_{f}^{2} & -\frac{1}{4}\mu\omega_{f}^{2} & 0\\
-\frac{1}{4}\mu\omega_{f}^{2} & \frac{1}{2}\mu\omega_{i}^{2}+\frac{1}{4}\mu\omega_{f}^{2} & 0\\
0 & 0 & \mu\omega_{i}^{2}+\mu\omega_{f}^{2}
\end{array}\right).
\end{equation}
Note that $A_{23}=0$, the matrix $A$ can be transformed into a diagonal matrix
\begin{equation}
A^{\prime}=\left(\begin{array}{ccc}
A_{11}-\frac{A_{12}^{2}}{A_{22}} & 0 & 0\\
0 & A_{22} & 0\\
0 & 0 & A_{33}
\end{array}\right),
\end{equation}
through the coordinate transformations,  $\boldsymbol{\xi}_{1}^{\prime}=\boldsymbol{\xi}_{1}$, $\boldsymbol{\xi}_{2}^{\prime}=\frac{-A^{\prime}_{12}}{A^{\prime}_{22}}\boldsymbol{\xi}_{1}
+\frac{-A^{\prime}_{12}}{A^{\prime}_{22}}\boldsymbol{\xi}_{2}$, $\boldsymbol{\xi}_{3}^{\prime}=\boldsymbol{\xi}_{3}$. On the other hand, in the calculations, the
plane wave should be expanded by
\begin{equation}
\begin{array}{ccc}
e^{i\boldsymbol{P}\cdot\boldsymbol{r}} & = & \boldsymbol{\sum}_{l=0}^{\infty}\sqrt{4\pi\left(2l+1\right)}i^{l}j_{l}\left(Pr\right)Y_{l0}\left(\hat{\boldsymbol{r}}\right)\end{array},
\end{equation}
where we let the momentum $P$ along the $z$ direction.
With the above steps, one can obtain the integration of the spatial part.

\begin{table*}[htbp]
\centering{} \caption{\label{squareradius}The harmonic oscillator parameters for the $T_{(4Q)}$ state. }\label{t4q}.
\scalebox{1.0}{%
\begin{tabular}{ccc|ccc||ccc|ccc}
\hline \hline
 & State  & $\beta$  & State  & $\beta$  &  &  & State  & $\beta$  & State  & $\beta$  & \tabularnewline
\hline
 & $T_{(cc\bar{c}\bar{c})0^{++}}(1S)(6455)$  & 481  & $T_{(cc\bar{c}\bar{c})1^{+-}}(1S)(6500)$  & 493  &  &  & $T_{(bb\bar{b}\bar{b})0^{++}}(1S)(19306)$  & 897 & $T_{(bb\bar{b}\bar{b})1^{+-}}(1S)(19329)$  & 897 & \tabularnewline
 & $T_{(cc\bar{c}\bar{c})0^{++}}(1S)(6550)$  & 493  & $T_{(cc\bar{c}\bar{c})2^{++}}(1S)(6524)$  & 481  &  &  & $T_{(bb\bar{b}\bar{b})0^{++}}(1S)(19355)$  & 897 & $T_{(bb\bar{b}\bar{b})2^{++}}(1S)(19341)$  & 897 & \tabularnewline
%\hline
 & $T_{(cc\bar{c}\bar{c})0^{+-}}(2S)(6993)$  & 403  & $T_{(cc\bar{c}\bar{c})1^{+-}}(2S)(6919)$  & 420  &  &  & $T_{(bb\bar{b}\bar{b})0^{+-}}(2S)(19789)$  & 680 & $T_{(bb\bar{b}\bar{b})1^{+-}}(2S)(19722)$  & 704 & \tabularnewline
 & $T_{(cc\bar{c}\bar{c})0^{+-}}(2S)(7020)$  & 420  & $T_{(cc\bar{c}\bar{c})1^{+-}}(2S)(7021)$  & 411  &  &  & $T_{(bb\bar{b}\bar{b})0^{+-}}(2S)(19841)$  & 704 & $T_{(bb\bar{b}\bar{b})1^{+-}}(2S)(19813)$  & 704 & \tabularnewline
 & $T_{(cc\bar{c}\bar{c})0^{++}}(2S)(6908)$  & 411  & $T_{(cc\bar{c}\bar{c})1^{++}}(2S)(7009)$  & 420  &  &  & $T_{(bb\bar{b}\bar{b})0^{++}}(2S)(19719)$  & 704 & $T_{(bb\bar{b}\bar{b})1^{++}}(2S)(19792)$  & 704 & \tabularnewline
 & $T_{(cc\bar{c}\bar{c})0^{++}}(2S)(6957)$  & 420  & $T_{(cc\bar{c}\bar{c})2^{+-}}(2S)(7017)$  & 411  &  &  & $T_{(bb\bar{b}\bar{b})0^{++}}(2S)(19767)$  & 730 & $T_{(bb\bar{b}\bar{b})2^{+-}}(2S)(19795)$  & 704 & \tabularnewline
 & $T_{(cc\bar{c}\bar{c})0^{++}}(2S)(7018)$  & 420  & $T_{(cc\bar{c}\bar{c})2^{++}}(2S)(6927)$  & 411  &  &  & $T_{(bb\bar{b}\bar{b})0^{++}}(2S)(19811)$  & 704 & $T_{(bb\bar{b}\bar{b})2^{++}}(2S)(19726)$  & 704 & \tabularnewline
 & $T_{(cc\bar{c}\bar{c})0^{++}}(2S)(7185)$  & 411  & $T_{(cc\bar{c}\bar{c})2^{++}}(2S)(7032)$  & 411  &  &  & $T_{(bb\bar{b}\bar{b})0^{++}}(2S)(19976)$  & 704 & $T_{(bb\bar{b}\bar{b})2^{++}}(2S)(19816)$  & 704 & \tabularnewline
%\hline
 & $T_{(cc\bar{c}\bar{c})0^{--}}(1P)(6651)$  & 438  & $T_{(cc\bar{c}\bar{c})1^{--}}(1P)(6904)$  & 438  &  &  & $T_{(bb\bar{b}\bar{b})0^{--}}(1P)(19485)$  & 789 & $T_{(bb\bar{b}\bar{b})1^{--}}(1P)(19748)$  & 789 & \tabularnewline
 & $T_{(cc\bar{c}\bar{c})0^{--}}(1P)(6926)$  & 438  & $T_{(cc\bar{c}\bar{c})1^{--}}(1P)(6993)$  & 438  &  &  & $T_{(bb\bar{b}\bar{b})0^{--}}(1P)(19756)$  & 759 & $T_{(bb\bar{b}\bar{b})1^{--}}(1P)(19795)$  & 759 & \tabularnewline
 & $T_{(cc\bar{c}\bar{c})0^{-+}}(1P)(6676)$  & 438  & $T_{(cc\bar{c}\bar{c})1^{-+}}(1P)(6675)$  & 438  &  &  & $T_{(bb\bar{b}\bar{b})0^{-+}}(1P)(19500)$  & 789 & $T_{(bb\bar{b}\bar{b})1^{-+}}(1P)(19496)$  & 789 & \tabularnewline
 & $T_{(cc\bar{c}\bar{c})0^{-+}}(1P)(6748)$  & 438  & $T_{(cc\bar{c}\bar{c})1^{-+}}(1P)(6768)$  & 438  &  &  & $T_{(bb\bar{b}\bar{b})0^{-+}}(1P)(19595)$  & 759 & $T_{(bb\bar{b}\bar{b})1^{-+}}(1P)(19603)$  & 759 & \tabularnewline
 & $T_{(cc\bar{c}\bar{c})0^{-+}}(1P)(6897)$  & 438  & $T_{(cc\bar{c}\bar{c})1^{-+}}(1P)(6910)$  & 438  &  &  & $T_{(bb\bar{b}\bar{b})0^{-+}}(1P)(19739)$  & 789 & $T_{(bb\bar{b}\bar{b})1^{-+}}(1P)(19748)$  & 789 & \tabularnewline
 & $T_{(cc\bar{c}\bar{c})1^{--}}(1P)(6636)$  & 438  & $T_{(cc\bar{c}\bar{c})2^{--}}(1P)(6630)$  & 438  &  &  & $T_{(bb\bar{b}\bar{b})1^{--}}(1P)(19479)$  & 789 & $T_{(bb\bar{b}\bar{b})2^{--}}(1P)(19476)$  & 789 & \tabularnewline
 & $T_{(cc\bar{c}\bar{c})1^{--}}(1P)(6750)$  & 438  & $T_{(cc\bar{c}\bar{c})2^{--}}(1P)(6780)$  & 438  &  &  & $T_{(bb\bar{b}\bar{b})1^{--}}(1P)(19597)$  & 759 & $T_{(bb\bar{b}\bar{b})2^{--}}(1P)(19608)$  & 759 & \tabularnewline
 & $T_{(cc\bar{c}\bar{c})1^{--}}(1P)(6768)$  & 438  & $T_{(cc\bar{c}\bar{c})2^{--}}(1P)(6955)$  & 438  &  &  & $T_{(bb\bar{b}\bar{b})1^{--}}(1P)(19603)$  & 759 & $T_{(bb\bar{b}\bar{b})2^{--}}(1P)(19767)$  & 759 & \tabularnewline
 & $T_{(cc\bar{c}\bar{c})2^{-+}}(1P)(6667)$  & 438  & $T_{(cc\bar{c}\bar{c})2^{-+}}(1P)(6783)$  & 438  &  &  & $T_{(bb\bar{b}\bar{b})2^{-+}}(1P)(19492)$  & 789 & $T_{(bb\bar{b}\bar{b})2^{-+}}(1P)(19609)$  & 759 & \tabularnewline
 & $T_{(cc\bar{c}\bar{c})2^{-+}}(1P)(6928)$  & 438  & $T_{(cc\bar{c}\bar{c})3^{--}}(1P)(6801)$  & 438  &  &  & $T_{(bb\bar{b}\bar{b})2^{-+}}(1P)(19756)$  & 789  & $T_{(bb\bar{b}\bar{b})3^{--}}(1P)(19617)$  & 759 & \tabularnewline
\hline \hline
\end{tabular}}
\end{table*}

%\subsection{Components of color configurations and root mean square radii for the physical tetraquark states  }

%\section{}

The components of different color configurations
for the physical $T_{(4c)}$ and $T_{(4b)}$ states are given in the Tables~\ref{rmsc} and~\ref{rmsb}, respectively.
To know the spatial size of the tetraquark states,
we also calculate the root mean square radius, our results are listed in Tables~\ref{rmsc} and~\ref{rmsb} as well.

\begin{table*}[htp]
\centering{}
\caption{The components of different color configurations and the root mean
square radius (fm) for each physical $T_{(4c)}$ states\label{rmsc},
where $\ensuremath{\boldsymbol{r}_{12-34}\equiv\left(\mathbf{r}_{1}+\mathbf{r}_{2}\right)/2-\left(\mathbf{r}_{3}+\mathbf{r}_{4}\right)/2}$,
$\ensuremath{\boldsymbol{r}_{13-24}\equiv\left(\mathbf{r}_{1}+\mathbf{r}_{3}\right)/2-\left(\mathbf{r}_{2}+\mathbf{r}_{4}\right)/2}$.}
\begin{tabular}{cr@{\extracolsep{0pt}.}lr@{\extracolsep{0pt}.}lr@{\extracolsep{0pt}.}lr@{\extracolsep{0pt}.}lcccc}
\hline\hline
~~~~~~~~~~~~~~State  ~~~~~~~~~~~~~~& \multicolumn{2}{c}{$\left|6\bar{6}\right\rangle_c $}~~~~~~~~ & \multicolumn{2}{c}{$\left|\bar{3}3\right\rangle_c $}~~~~~~~~~ & \multicolumn{2}{c}{$\left|11\right\rangle_c $} ~~~~~~~~~& \multicolumn{2}{c}{$\left|88\right\rangle_c $}~~~~~~~ & $\sqrt{\langle r_{12}^{2}\rangle}$ ~~~~~~~ & $\sqrt{\langle r_{12-34}^{2}\rangle}$  ~~~~~~~& $\sqrt{\langle r_{13}^{2}\rangle}$  ~~~~~~~& $\sqrt{\langle r_{13-24}^{2}\rangle}$ \tabularnewline
\hline
$T_{(4c)0^{++}}(6455)(1S)$  & 33&9$\%$  & 66&1$\%$  & 44&6$\%$  & 55&4$\%$  & 0.49  & 0.35  & 0.49  & 0.35\tabularnewline
$T_{(4c)0^{++}}(6550)(1S)$  & 66&1$\%$  & 33&9$\%$  & 55&4$\%$  & 44&6$\%$  & 0.50  & 0.35  & 0.50  & 0.35\tabularnewline
$T_{(4c)1^{+-}}(6500)(1S)$  & 0&0$\%$  & 100&0$\%$  & 33&3$\%$  & 66&7$\%$  & 0.50  & 0.35  & 0.50  & 0.35\tabularnewline
$T_{(4c)2^{++}}(6524)(1S)$  & 0&0$\%$  & 100&0$\%$  & 33&3$\%$  & 66&7$\%$  & 0.51  & 0.36  & 0.51  & 0.36\tabularnewline
$T_{(4c)0^{--}}(6651)(1P)$  & 64&0$\%$  & 36&0$\%$  & 54&7$\%$  & 45&3$\%$  & 0.63  & 0.39  & 0.59  & 0.45\tabularnewline
$T_{(4c)0^{--}}(6926)(1P)$  & 36&0$\%$  & 64&0$\%$  & 45&3$\%$  & 54&7$\%$  & 0.63  & 0.39  & 0.59  & 0.45\tabularnewline
$T_{(4c)0^{-+}}(6681)(1P)$  & 67&5$\%$  & 32&5$\%$  & 55&8$\%$  & 44&2$\%$  & 0.61  & 0.39  & 0.59  & 0.43\tabularnewline
$T_{(4c)0^{-+}}(6749)(1P)$  & 2&0$\%$  & 98&0$\%$  & 34&0$\%$  & 66&0$\%$  & 0.55  & 0.47  & 0.61  & 0.39\tabularnewline
$T_{(4c)0^{-+}}(6891)(1P)$  & 30&1$\%$  & 69&9$\%$  & 43&4$\%$  & 56&6$\%$  & 0.62  & 0.39  & 0.59  & 0.44\tabularnewline
$T_{(4c)1^{--}}(6636)(1P)$  & 67&8$\%$  & 32&2$\%$  & 55&9$\%$  & 44&1$\%$  & 0.63  & 0.39  & 0.59  & 0.44\tabularnewline
$T_{(4c)1^{--}}(6750)(1P)$  & 0&4$\%$  & 99&6$\%$  & 33&5$\%$  & 66&5$\%$  & 0.54  & 0.48  & 0.61  & 0.38\tabularnewline
$T_{(4c)1^{--}}(6768)(1P)$  & 1&0$\%$  & 99&0$\%$  & 33&7$\%$  & 66&3$\%$  & 0.55  & 0.50  & 0.63  & 0.39\tabularnewline
$T_{(4c)1^{--}}(6904)(1P)$  & 48&4$\%$  & 51&6$\%$  & 49&5$\%$  & 50&5$\%$  & 0.60  & 0.42  & 0.60  & 0.43\tabularnewline
$T_{(4c)1^{--}}(6993)(1P)$  & 83&5$\%$  & 16&5$\%$  & 61&2$\%$  & 38&8$\%$  & 0.57  & 0.47  & 0.62  & 0.4\tabularnewline
$T_{(4c)1^{-+}}(6676)(1P)$  & 68&0$\%$  & 32&0$\%$  & 56&0$\%$  & 44&0$\%$  & 0.63  & 0.39  & 0.59  & 0.45\tabularnewline
$T_{(4c)1^{-+}}(6769)(1P)$  & 0&0$\%$  & 100&0$\%$  & 33&3$\%$  & 66&7$\%$  & 0.55  & 0.50  & 0.63  & 0.39\tabularnewline
$T_{(4c)1^{-+}}(6908)(1P)$  & 32&5$\%$  & 67&5$\%$  & 44&2$\%$  & 55&8$\%$  & 0.62  & 0.38  & 0.59  & 0.44\tabularnewline
$T_{(4c)2^{--}}(6630)(1P)$  & 64&5$\%$  & 35&5$\%$  & 54&8$\%$  & 45&2$\%$  & 0.64  & 0.39  & 0.60  & 0.45\tabularnewline
$T_{(4c)2^{--}}(6780)(1P)$  & 0&0$\%$  & 100&0$\%$  & 33&3$\%$  & 66&7$\%$  & 0.55  & 0.50  & 0.63  & 0.39\tabularnewline
$T_{(4c)2^{--}}(6955)(1P)$  & 35&6$\%$  & 64&4$\%$  & 45&2$\%$  & 54&8$\%$  & 0.63  & 0.39  & 0.59  & 0.45\tabularnewline
$T_{(4c)2^{-+}}(6667)(1P)$  & 67&2$\%$  & 32&8$\%$  & 55&7$\%$  & 44&3$\%$  & 0.63  & 0.39  & 0.59  & 0.45\tabularnewline
$T_{(4c)2^{-+}}(6783)(1P)$  & 0&0$\%$  & 100&0$\%$  & 33&3$\%$  & 66&7$\%$  & 0.55  & 0.50  & 0.64  & 0.39\tabularnewline
$T_{(4c)2^{-+}}(6928)(1P)$  & 33&7$\%$  & 66&3$\%$  & 44&6$\%$  & 55&4$\%$  & 0.63  & 0.39  & 0.59  & 0.45\tabularnewline
$T_{(4c)3^{--}}(6801)(1P)$  & 0&0$\%$  & 100&0$\%$  & 33&3$\%$  & 66&7$\%$  & 0.56  & 0.51  & 0.64  & 0.39\tabularnewline
$T_{(4c)0^{+-}}(6993)(2S)$  & 43&6$\%$  & 56&4$\%$  & 47&9$\%$  & 52&1$\%$  & 0.77  & 0.43  & 0.69  & 0.54\tabularnewline
$T_{(4c)0^{+-}}(7020)(2S)$  & 56&4$\%$  & 43&6$\%$  & 52&1$\%$  & 47&9$\%$  & 0.78  & 0.42  & 0.69  & 0.55\tabularnewline
$T_{(4c)0^{++}}(6908)(2S)$  & 12&5$\%$  & 87&5$\%$  & 37&5$\%$  & 62&5$\%$  & 0.45  & 0.66  & 0.73  & 0.32\tabularnewline
$T_{(4c)0^{++}}(6957)(2S)$  & 83&0$\%$  & 17&0$\%$  & 61&0$\%$  & 39&0$\%$  & 0.76  & 0.42  & 0.69  & 0.54\tabularnewline
$T_{(4c)0^{++}}(7018)(2S)$  & 4&4$\%$  & 95&6$\%$  & 34&8$\%$  & 65&2$\%$  & 0.56  & 0.56  & 0.69  & 0.40\tabularnewline
$T_{(4c)0^{++}}(7185)(2S)$  & 99&7$\%$  & 0&3$\%$  & 66&6$\%$  & 33&4$\%$  & 0.74  & 0.33  & 0.62  & 0.52\tabularnewline
$T_{(4c)1^{+-}}(6919)(2S)$  & 0&0$\%$  & 100&0$\%$  & 33&3$\%$  & 66&7$\%$  & 0.52  & 0.72  & 0.82  & 0.37\tabularnewline
$T_{(4c)1^{+-}}(7021)(2S)$  & 0&0$\%$  & 100&0$\%$  & 33&3$\%$  & 66&7$\%$  & 0.69  & 0.36  & 0.61  & 0.49\tabularnewline
$T_{(4c)1^{++}}(7009)(2S)$  & 0&0$\%$  & 100&0$\%$  & 33&3$\%$  & 66&7$\%$  & 0.73  & 0.45  & 0.68  & 0.51\tabularnewline
$T_{(4c)2^{+-}}(7017)(2S)$  & 0&0$\%$  & 100&0$\%$  & 33&3$\%$  & 66&7$\%$  & 0.73  & 0.45  & 0.68  & 0.52\tabularnewline
$T_{(4c)2^{++}}(6927)(2S)$  & 0&0$\%$  & 100&0$\%$  & 33&3$\%$  & 66&7$\%$  & 0.47  & 0.68  & 0.76  & 0.33\tabularnewline
$T_{(4c)2^{++}}(7032)(2S)$  & 0&0$\%$  & 100&0$\%$  & 33&3$\%$  & 66&7$\%$  & 0.79  & 0.45  & 0.72  & 0.56\tabularnewline
\hline\hline
\end{tabular}
\end{table*}

\begin{table*}[htp]
\centering{}\caption{The components of different color configurations and the root mean
square radius (fm) for each physical $T_{(4b)}$ states\label{rmsc},
where $\ensuremath{\boldsymbol{r}_{12-34}\equiv\left(\mathbf{r}_{1}+\mathbf{r}_{2}\right)/2-\left(\mathbf{r}_{3}+\mathbf{r}_{4}\right)/2}$,
$\ensuremath{\boldsymbol{r}_{13-24}\equiv\left(\mathbf{r}_{1}+\mathbf{r}_{3}\right)/2-\left(\mathbf{r}_{2}+\mathbf{r}_{4}\right)/2}$\label{rmsb}.}
\begin{tabular}{ccccccccc}
\hline\hline
~~~~~~State~~~~~~  & ~~~~~~$\left|6\bar{6}\right\rangle_c $~~~~~~  & ~~~~~~$\left|\bar{3}3\right\rangle_c $ ~~~~~~ & ~~~~~~$\left|11\right\rangle_c $ ~~~~ & ~~~~$\left|88\right\rangle_c $ ~~~~ & ~~~~$\sqrt{\langle r_{12}^{2}\rangle}$ ~~~~& ~~~~$\sqrt{\langle r_{12-34}^{2}\rangle}$  ~~~~& ~~~~$\sqrt{\langle r_{13}^{2}\rangle}$~~~~  & ~~~~$\sqrt{\langle r_{13-24}^{2}\rangle}$ \tabularnewline
\hline
$T_{(4b)0^{++}}(19306)(1S)$  & 33.9$\%$  & 66.1$\%$  & 44.6$\%$  & 55.4$\%$  & 0.27  & 0.19  & 0.27  & 0.19\tabularnewline
$T_{(4b)0^{++}}(19355)(1S)$  & 66.1$\%$  & 33.9$\%$  & 55.4$\%$  & 44.6$\%$  & 0.27  & 0.19  & 0.27  & 0.19\tabularnewline
$T_{(4b)1^{+-}}(19329)(1S)$  & 0.0$\%$  & 100.0$\%$  & 33.3$\%$  & 66.7$\%$  & 0.27  & 0.19  & 0.27  & 0.19\tabularnewline
$T_{(4b)2^{++}}(19341)(1S)$  & 0.0$\%$  & 100.0$\%$  & 33.3$\%$  & 66.7$\%$  & 0.27  & 0.19  & 0.27  & 0.19\tabularnewline
$T_{(4b)0^{--}}(19485)(1P)$  & 65.3$\%$  & 34.7$\%$  & 55.1$\%$  & 44.9$\%$  & 0.36  & 0.22  & 0.34  & 0.25\tabularnewline
$T_{(4b)0^{--}}(19756)(1P)$  & 34.7$\%$  & 65.3$\%$  & 44.9$\%$  & 55.1$\%$  & 0.36  & 0.22  & 0.34  & 0.26\tabularnewline
$T_{(4b)0^{-+}}(19500)(1P)$  & 68.0$\%$  & 32.0$\%$  & 56.0$\%$  & 44.0$\%$  & 0.35  & 0.22  & 0.33  & 0.25\tabularnewline
$T_{(4b)0^{-+}}(19595)(1P)$  & 0.0$\%$  & 100.0$\%$  & 33.3$\%$  & 66.7$\%$  & 0.31  & 0.28  & 0.36  & 0.22\tabularnewline
$T_{(4b)0^{-+}}(19739)(1P)$  & 32.5$\%$  & 67.5$\%$  & 44.2$\%$  & 55.8$\%$  & 0.36  & 0.22  & 0.33  & 0.25\tabularnewline
$T_{(4b)1^{--}}(19479)(1P)$  & 67.1$\%$  & 32.9$\%$  & 55.7$\%$  & 44.3$\%$  & 0.36  & 0.22  & 0.34  & 0.26\tabularnewline
$T_{(4b)1^{--}}(19597)(1P)$  & 91.9$\%$  & 8.1$\%$  & 64.0$\%$  & 36.0$\%$  & 0.31  & 0.28  & 0.36  & 0.22\tabularnewline
$T_{(4b)1^{--}}(19603)(1P)$  & 0.0$\%$  & 100.0$\%$  & 33.3$\%$  & 66.7$\%$  & 0.31  & 0.28  & 0.36  & 0.22\tabularnewline
$T_{(4b)1^{--}}(19748)(1P)$  & 0.2$\%$  & 99.8$\%$  & 33.4$\%$  & 66.6$\%$  & 0.36  & 0.23  & 0.35  & 0.25\tabularnewline
$T_{(4b)1^{--}}(19795)(1P)$  & 41.3$\%$  & 58.7$\%$  & 47.1$\%$  & 52.9$\%$  & 0.33  & 0.29  & 0.37  & 0.23\tabularnewline
$T_{(4b)1^{-+}}(19496)(1P)$  & 67.3$\%$  & 32.7$\%$  & 55.8$\%$  & 44.2$\%$  & 0.36  & 0.22  & 0.34  & 0.25\tabularnewline
$T_{(4b)1^{-+}}(19603)(1P)$  & 0.0$\%$  & 100.0$\%$  & 33.3$\%$  & 66.7$\%$  & 0.31  & 0.28  & 0.36  & 0.22\tabularnewline
$T_{(4b)1^{-+}}(19748)(1P)$  & 32.5$\%$  & 67.5$\%$  & 44.2$\%$  & 55.8$\%$  & 0.36  & 0.22  & 0.34  & 0.26\tabularnewline
$T_{(4b)2^{--}}(19476)(1P)$  & 65.3$\%$  & 34.7$\%$  & 55.1$\%$  & 44.9$\%$  & 0.36  & 0.22  & 0.34  & 0.25\tabularnewline
$T_{(4b)2^{--}}(19608)(1P)$  & 0.0$\%$  & 100.0$\%$  & 33.3$\%$  & 66.7$\%$  & 0.31  & 0.29  & 0.36  & 0.22\tabularnewline
$T_{(4b)2^{--}}(19767)(1P)$  & 34.6$\%$  & 65.4$\%$  & 44.9$\%$  & 55.1$\%$  & 0.36  & 0.22  & 0.34  & 0.26\tabularnewline
$T_{(4b)2^{-+}}(19492)(1P)$  & 66.7$\%$  & 33.3$\%$  & 55.6$\%$  & 44.4$\%$  & 0.36  & 0.22  & 0.34  & 0.25\tabularnewline
$T_{(4b)2^{-+}}(19609)(1P)$  & 0.0$\%$  & 100.0$\%$  & 33.3$\%$  & 66.7$\%$  & 0.31  & 0.29  & 0.36  & 0.22\tabularnewline
$T_{(4b)2^{-+}}(19756)(1P)$  & 33.3$\%$  & 66.7$\%$  & 44.4$\%$  & 55.6$\%$  & 0.36  & 0.22  & 0.34  & 0.26\tabularnewline
$T_{(4b)3^{--}}(19617)(1P)$  & 0.0$\%$  & 100.0$\%$  & 33.3$\%$  & 66.7$\%$  & 0.32  & 0.29  & 0.36  & 0.22\tabularnewline
$T_{(4b)0^{+-}}(19789)(2S)$  & 1.0$\%$  & 99.0$\%$  & 33.7$\%$  & 66.3$\%$  & 0.43  & 0.28  & 0.42  & 0.31\tabularnewline
$T_{(4b)0^{+-}}(19841)(2S)$  & 99.0$\%$  & 1.0$\%$  & 66.3$\%$  & 33.7$\%$  & 0.48  & 0.24  & 0.42  & 0.34\tabularnewline
$T_{(4b)0^{++}}(19719)(2S)$  & 2.4$\%$  & 97.6$\%$  & 34.1$\%$  & 65.9$\%$  & 0.41  & 0.20  & 0.35  & 0.29\tabularnewline
$T_{(4b)0^{++}}(19767)(2S)$  & 1.0$\%$  & 99.0$\%$  & 33.7$\%$  & 66.3$\%$  & 0.28  & 0.43  & 0.48  & 0.20\tabularnewline
$T_{(4b)0^{++}}(19811)(2S)$  & 96.2$\%$  & 3.8$\%$  & 65.4$\%$  & 34.6$\%$  & 0.32  & 0.36  & 0.43  & 0.22\tabularnewline
$T_{(4b)0^{++}}(19976)(2S)$  & 100.0$\%$  & 0.0$\%$  & 66.7$\%$  & 33.3$\%$  & 0.39  & 0.16  & 0.31  & 0.28\tabularnewline
$T_{(4b)1^{+-}}(19722)(2S)$  & 0.0$\%$  & 100.0$\%$  & 33.3$\%$  & 66.7$\%$  & 0.40  & 0.20  & 0.35  & 0.28\tabularnewline
$T_{(4b)1^{+-}}(19813)(2S)$  & 0.0$\%$  & 100.0$\%$  & 33.3$\%$  & 66.7$\%$  & 0.30  & 0.44  & 0.49  & 0.21\tabularnewline
$T_{(4b)1^{++}}(19792)(2S)$  & 0.0$\%$  & 100.0$\%$  & 33.3$\%$  & 66.7$\%$  & 0.43  & 0.28  & 0.41  & 0.30\tabularnewline
$T_{(4b)2^{+-}}(19795)(2S)$  & 0.0$\%$  & 100.0$\%$  & 33.3$\%$  & 66.7$\%$  & 0.43  & 0.28  & 0.42  & 0.31\tabularnewline
$T_{(4b)2^{++}}(19726)(2S)$  & 0.0$\%$  & 100.0$\%$  & 33.3$\%$  & 66.7$\%$  & 0.38  & 0.20  & 0.34  & 0.27\tabularnewline
$T_{(4b)2^{++}}(19816)(2S)$  & 0.0$\%$  & 100.0$\%$  & 33.3$\%$  & 66.7$\%$  & 0.31  & 0.43  & 0.49  & 0.22\tabularnewline
\hline\hline
\end{tabular}
\end{table*}

\end{document}